\documentclass[lettersize,journal]{IEEEtran}
\usepackage{amsmath,amsfonts}
\usepackage{algorithmic}
\usepackage{algorithm}
\usepackage{array}
\usepackage[caption=false,font=normalsize,labelfont=sf,textfont=sf]{subfig}
\usepackage{textcomp}
\usepackage{stfloats}
\usepackage{url}
\usepackage{verbatim}
\usepackage{graphicx}
\usepackage{cite}
\usepackage{booktabs,ragged2e}
\usepackage[flushleft]{threeparttable}
\usepackage{tabularx}
\usepackage{tipa}
\usepackage{hyperref}
\usepackage{enumitem}

\begin{document}

\title{Topological data analysis of human vowels: \\ Persistent homologies across  representation spaces}
\author{
    \IEEEauthorblockN{Guillem Bonafos\IEEEauthorrefmark{1}\IEEEauthorrefmark{2}\IEEEauthorrefmark{3}\IEEEauthorrefmark{4}, Pierre Pudlo\IEEEauthorrefmark{1}\IEEEauthorrefmark{3}, Jean-Marc Freyermuth\IEEEauthorrefmark{1}\IEEEauthorrefmark{3}\IEEEauthorrefmark{5}, Samuel Tronçon\IEEEauthorrefmark{4}, Arnaud Rey\IEEEauthorrefmark{2}\IEEEauthorrefmark{3}}\\
    \smallskip
    \IEEEauthorblockA{\IEEEauthorrefmark{1}Aix Marseille Univ, CNRS, I2M, Marseille, France}\\
    \IEEEauthorblockA{\IEEEauthorrefmark{2}Aix Marseille Univ, CNRS, LPC, Marseille, France}\\
    \IEEEauthorblockA{\IEEEauthorrefmark{3}Aix Marseille Univ, ILCB, Aix-en-Provence, France}\\
    \IEEEauthorblockA{\IEEEauthorrefmark{4}Résurgences R\&D, Arles, France}
\thanks{\IEEEauthorrefmark{5}This publication is based upon
work supported by King Abdullah University of Science and Technology Research Funding (KRF) under Award No.
ORFS-2022-CRG11-5025.2}
}



\maketitle

\begin{abstract}

Topological Data Analysis (TDA) has been successfully used for various tasks in signal/image processing, from visualization to supervised/unsupervised classification. Often, topological characteristics are obtained from persistent homology theory. The standard TDA pipeline starts from the raw signal data or a representation of it. Then, it consists in building a multiscale topological structure on the top of the data using a pre-specified filtration, and finally to compute the topological signature to be further exploited. The commonly used topological signature is a persistent diagram (or transformations of it). Current research discusses the consequences of the many ways to exploit topological signatures, much less often the choice of the filtration, but to the best of our knowledge, the choice of the representation of a signal has not been the subject of any study yet. This paper attempts to provide some answers on the latter problem. To this end, we collected real audio data and built a comparative study to assess the quality of the discriminant information of the topological signatures extracted from three different representation spaces. Each audio signal is represented as i) an embedding of observed data in a higher dimensional space using Taken’s representation, ii) a spectrogram viewed as a surface in a 3D ambient space, iii) the set of spectrogram's zeroes. From vowel audio recordings, we use topological signature for three prediction problems: speaker gender, vowel type, and individual. We show that topologically-augmented random forest improves the Out-of-Bag Error (OOB) over solely based Mel-Frequency Cepstral Coefficients (MFCC) for the last two problems. Our results also suggest that the topological information extracted from different signal representations is complementary, and that spectrogram's zeros offers the best improvement for gender prediction. 

\end{abstract}


\section{Introduction}

\IEEEPARstart{T}{opological Data Analysis} (TDA) is a fast-growing research area that relies on deep mathematical foundations \cite{carlssonTopologyData2009, wassermanTopologicalDataAnalysis2018, chazalIntroductionTopologicalData2021}. It offers novel and potentially fruitful angles of analysis of digital audio signals. This innovative approach to data science is based on extracting information from the \emph{shape of data}.

TDA has already been applied to various signal processing problems \cite{barbarossaTopologicalSignalProcessing2020, barbarossaTopologicalSignalProcessing2020a}. It starts from the assumption that the data have a shape \cite{ferriWhyTopologyMachine2018} and it computes its persistent homologies, which provide a compact representation of its topological features. These are stable to perturbations of input data and independent of dimensions and coordinate systems. However, this shape strongly depends on the way a signal is represented (i.e., on the representation space). The purpose of this work is to study how the computation of persistent homologies depends on the chosen representation spaces, considering the specific problem of vowels categorization in human language. We study the impact of the representation space on the extracted topological information and determine if accessing higher-dimensional persistent homologies allows getting more discriminant information. We also discuss what is the best way to summarize the information contained in a persistence diagram for our specific classification tasks.

This article is organized as follows. First, we introduce the problem and its rationale. Second, we describe in a nutshell the theory and the processing pipeline of TDA. Third, we present the strategy for investigating the problem, the nature of the acquired dataset and our classification aims. Fourth, we report the main results that are extensively discussed in the last section. 

\section{The problem}

\subsection{Motivations for TDA}\label{part:motivations}

Topology is the branch of mathematics that deals with the qualitative geometric information of a space \cite{carlssonTopologyData2009}. The tools provided by algebraic topology allow us to capture the shape of the data \cite{zomorodianTopologyComputing2005}. The topological approach frees itself from the question of metrics and coordinates by studying the properties of a space through its connectivity. It has an interesting explanatory power thanks to its great potential for visualization, and 
the topological features have a discriminant power which makes TDA a particularly interesting candidate for the classification of natural signals. In this section, we give an overview of application of TDA in signal processing. Additional details on the theoretical foundations of TDA are given in Section \ref{part:theoretical_background}.

A central feature in TDA is the computation of persistent homologies \cite{otterRoadmapComputationPersistent2017}. Several pipelines have been proposed for the computation and use of such topological descriptors in data analysis. It typically consists in calculating persistent homologies on the input data (e.g. an audio signal) or a representation of it (e.g., its spectrogram), vectorizing persistence diagrams, and using these characteristics in a model \cite{chazalIntroductionTopologicalData2021, boissonnatTopologicalDataAnalysis2022b}. Persistence diagrams and some of their representations have been shown to be stable against noise \cite{cohen-steinerStabilityPersistenceDiagrams2007, chazalIntroductionTopologicalData2021, chazalStochasticConvergencePersistence2014}, meaning that small perturbations of the input data results in small changes in the persistent diagram. This stability makes the topological approach an excellent candidate for the description of natural signals. Altogether, the interesting perspectives offered by TDA to face Big Data challenges combined with the fast development of tools for the efficient computation of topological descriptors has led to a proliferation of studies demonstrating the added value of TDA in a variety of contexts.

For example, \cite{barbarossaTopologicalSignalProcessing2020, barbarossaTopologicalSignalProcessing2020a} demonstrate the usefulness of TDA for signal processing and for the study of signals on graphs. Topological tools are also useful in analyzing the shape of time series \cite{severskyTimeSeriesTopologicalData2016}, for object detection in images \cite{patrangenaruChallengesTopologicalObject2019} or in sound detection \cite{fireaizenAlarmSoundDetection2022}. Besides detection problems, topological descriptors are also useful for the classification of sound signals \cite{liuApplyingTopologicalPersistence2016} and musical signals \cite{bergomiHomologicalPersistenceTime2020}. It impacts multiple scientific domains alike biology, medicine, ecology  \cite{pereiraPersistentHomologyTime2015}, neurosciences, e.g., for fMRI \cite{salchMathematicsMedicinePractical2021} or EEG data \cite{nasrinBayesianTopologicalLearning2019, xuTopologicalDataAnalysis2021} for which TDA allows the construction of invariant signal descriptors. Topological features are complementary to more classical descriptors, and they allow capturing global and high-dimensional information useful for signal analysis. However, the impact of the representation space of the signal on the extracted topological information is, to our knowledge, still an unexplored question. 

\subsection{Representation space}\label{part:representation}

\begin{figure*}
\centering
  \subfloat[Wave\label{fig:representation_wav}]{\includegraphics[width=0.45\linewidth]{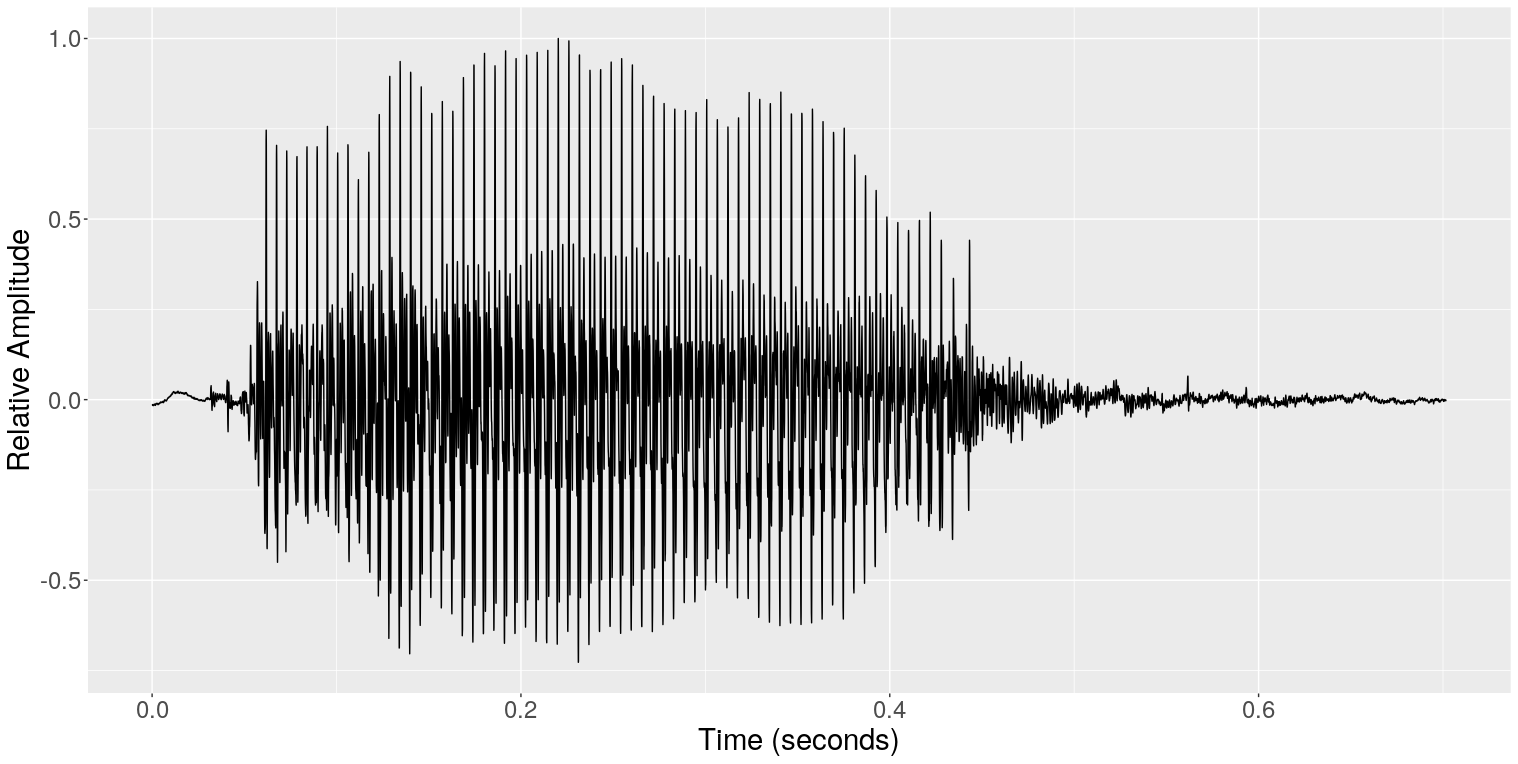}}
  \hfill
  \subfloat[Spectrogram's surface\label{fig:representation_spectro}]{\includegraphics[width=0.45\linewidth]{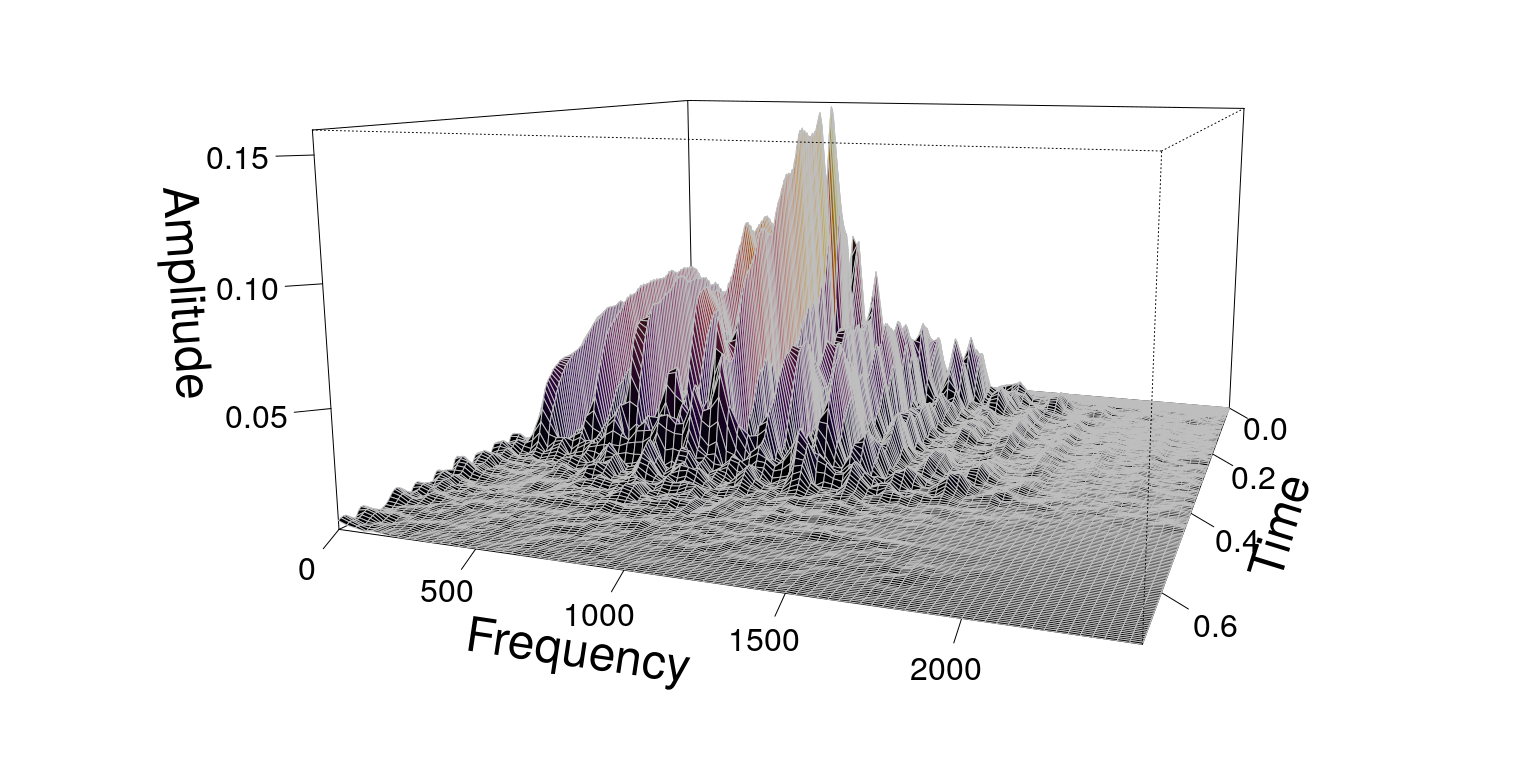}}
  \\
  \subfloat[Spectrogram's zeros\label{fig:representation_zeros}]{\includegraphics[width=0.45\linewidth]{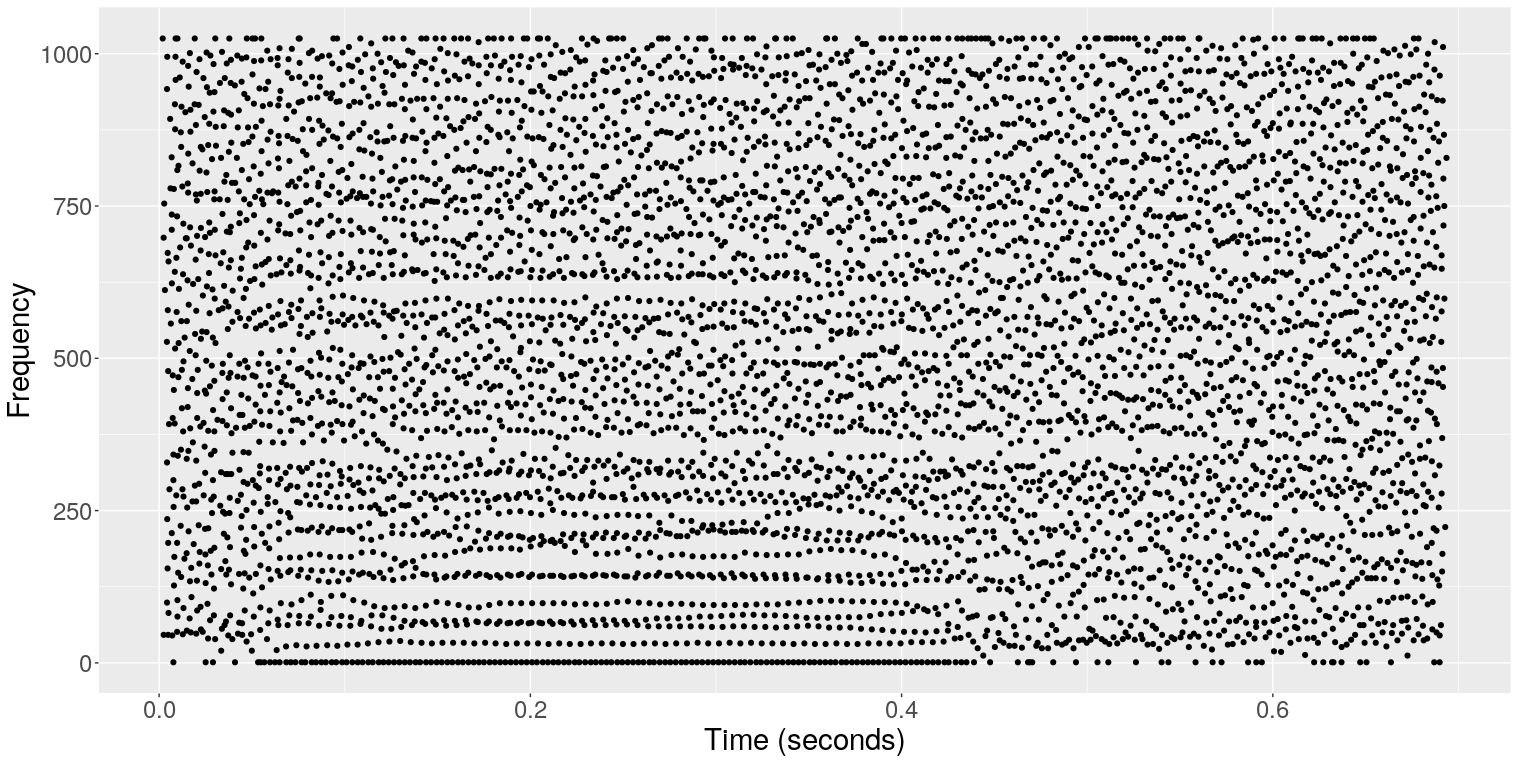}}
  \hfill
  \subfloat[Taken's embedding\label{fig:representation_taken}]{\includegraphics[width=0.45\linewidth]{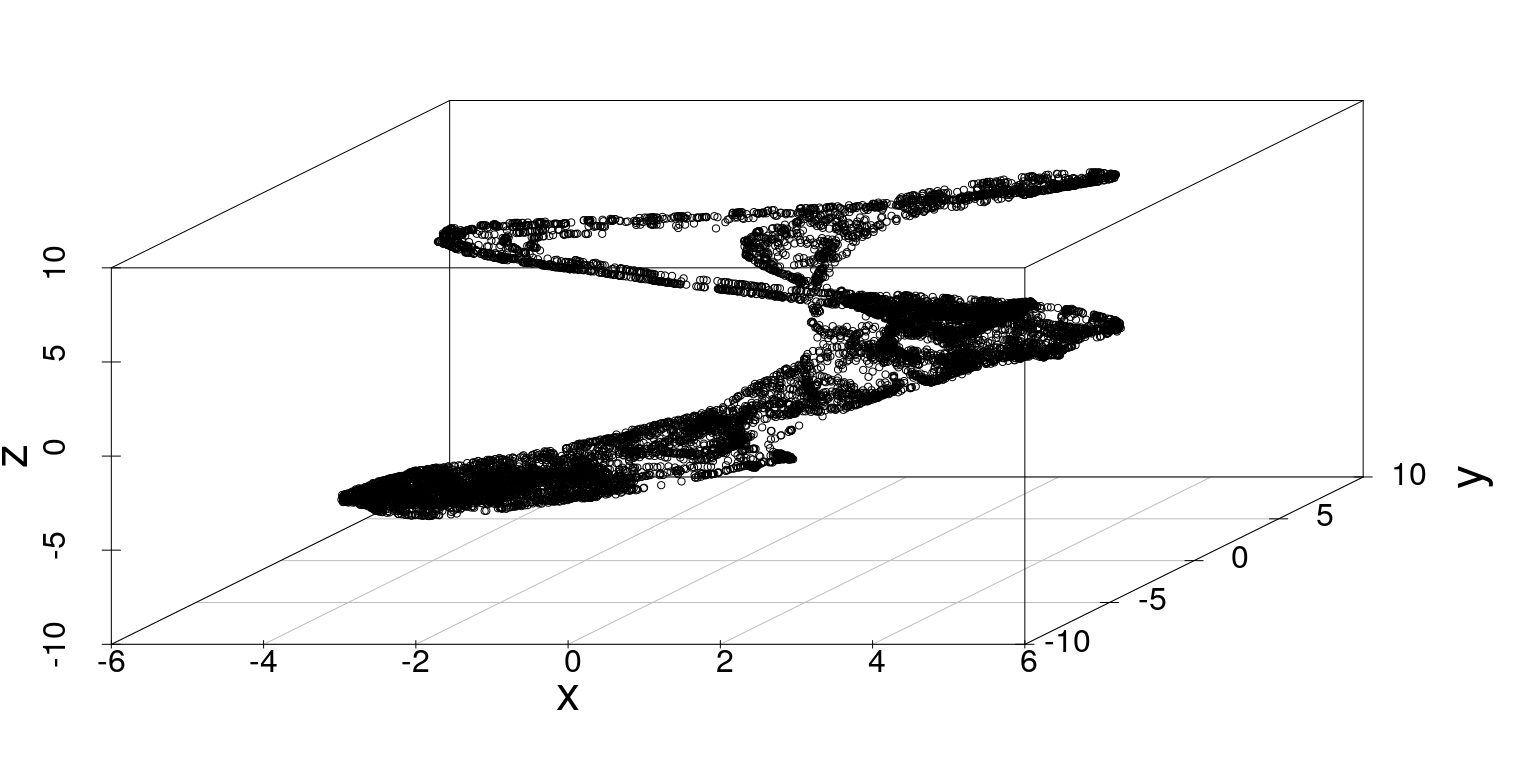}}
  \caption{Different representation of the same signal. \ref{fig:representation_wav} is the initial wave of the sound; \ref{fig:representation_spectro}) is the surface of its spectrogram; \ref{fig:representation_zeros} is the zeros of its spectrogram in the time-frequency plane; \ref{fig:representation_taken} is its Taken's embedding.}
  \label{fig:representation_signal} 
\end{figure*}

Our study focuses on physical signals, in particular, sound signals. A raw data signal can be represented as different \emph{data objects}, in different, here-called, representation spaces. Figure \ref{fig:representation_signal} shows the same raw signal in three different ways: as a spectrogram, viewed as a surface in 3D Euclidean ambient space, as the spectrogram's zeros, or as a point cloud using Taken's embedding.

Flandrin \cite{flandrinExplorationsTimeFrequencyAnalysis2018} refers to time-frequency analysis as the language of signal processing. Among the plethora of existing representation methods, the spectrogram is certainly the most widely used. Let us consider a sound signal $x\in L_1(\mathbb{R})\cap L_2(\mathbb{R})$, its spectrogram $S_x^{(h)}(t,\omega)$ is defined as:
\begin{equation}\label{eq:spectrogram}
    S_x^{(h)}(t,\omega)=|F_x^{(h)}(t,\omega)|^2,
\end{equation}
where $(t,\omega)\in \mathbb{R}^2$ are time and frequency variables, $h(t)\in\mathbb{R}$ is a window function and $F_x^{(h)}(t,\omega)$ is the Short-Time Fourier Transform (STFT):
\begin{equation}\label{eq:stft}
    F^{(h)}_x (t,\omega)=\int_{-\infty}^{+\infty}x(s)h(s-t)\exp\{-i\omega (s-t/2)\}ds.
\end{equation}

We can therefore represent the spectrogram of a one dimensional sound signal $x$, with time window function $h(\cdot)$, as a surface $\mathcal{S}_{x}^{(h)}$ in a 3D ambient Euclidean space where the dimensions are time, frequency and amplitude, $\mathcal{S}_{x}^{(h)}:=\left\{\left(t,\omega,S^{(h)}_x(t,\omega)\right)\vert (t,\omega)\in\mathbb{R}^2 \right\}\subset \mathbb{R}^3$ (see Figure \ref{fig:representation_spectro}). This representation has been proven effective in revealing geometrical structures, enhancing classification performances \cite{LevyClassificationofaudiosignals2022}. Another possibility is to rather consider the spectrogram's zeros as introduced in \cite{flandrinTimeFrequencyFiltering2015}. Choosing the window function to be Gaussian, i.e, $h(t)=\pi^{-1/4}\exp\left\{-t^2/4\right\}$, we readily show that the STFT can be rewritten as follows:
\begin{equation}
F_x^{(h)}(t,\omega)=\exp\left\{-\vert z\vert^2/4\right\}\mathcal{F}_x(z),
\end{equation}
where, $z=\omega + it$ and $\mathcal{F}_x(z)$ is the Bargmann transform of $x$. It is easy to see that it is an entire function of order $2$, which admits a Weierstrass-Hadamard form:
\begin{equation}
\mathcal{F}_x(z)\propto \prod_{n=1}^\infty\left(1-\frac{z}{z_n}\right)\exp\left\{\frac{z}{z_n}+\frac{1}{2}\left(\frac{z}{z_n}\right)^2\right\},
\end{equation}
where $\mathcal{Z}_x:=\left\{z_n=\omega_n+it_n\right\}_n$ is the set of zeroes of $\mathcal{F}_x(z)$. The spectrogram is therefore completely characterized by its zeroes, in other words, it can be represented by a point cloud in the time-frequency plane. \cite{flandrinExplorationsTimeFrequencyAnalysis2018} pioneered the idea of using topological characteristics for describing a spectrogram by visualizing the distribution of the edges of its zero-based Delaunay-triangulation. In the sequel, we build on this idea for extracting topological characteristics using the tools described hereafter in Section~(\ref{part:theoretical_background}). 

We introduce another representation of sound signals that is not related to time-frequency analysis. Considering a sound signal as a discrete-time digital audio recording (or time series) $\left\{x_1,\ldots,x_T\right\}$ and assuming it comes from a dynamic system, we can borrow the tools of dynamical system analysis to find out an informative representation. The Taken's theorem states that it is possible to obtain a representation of this time series that is topologically equivalent to the attractor of the system via a delay embedding, which contains useful information about the system \cite{takensDetectingStrangeAttractors1981a}. We thus transform the digital audio recording into a point cloud in a higher dimensional space, $\mathcal{P}_{D,x} = \left\{p_1,\ldots,p_m\right\}\subset\mathbb{R}^D$, each element $p_i \in \mathcal{P}_{D,x}$ is a vector of dimension $D$, constructed by taking a time delay $\tau$: 
\begin{equation}\label{eq:taken}
    p_i=(x_i, x_{i+\tau}, x_{i+2\tau},...,x_{i+(D-1)\tau})'.
\end{equation}

We therefore embed a one dimensional digital audio recording into a higher dimensional space to obtain a point cloud. It is necessary to estimate the two hyperparameters: $D$, the dimension of the space, and $\tau$, the time delay.  $D$ is estimated using the Cao's algorithm \cite{caoPracticalMethodDetermining1997}. $\tau$ is selected using the Average Mutual Information (AMI). The coordinates of the phase-space embedding must be independent enough (to avoid aggregation around the diagonal in the embedding). $\tau$ is then chosen to be the smallest value such that $AMI(\tau)<\frac{1}{e}$. An example of this point cloud representation is shown in Figure~\ref{fig:representation_taken}. 

\subsection{Same signal, different representations, different topological characteristics}

\begin{figure} 
    \centering
  \subfloat[Persistence diagram of the spectrogram's surface\label{fig:diagram_surf}]{\includegraphics[width=\linewidth]{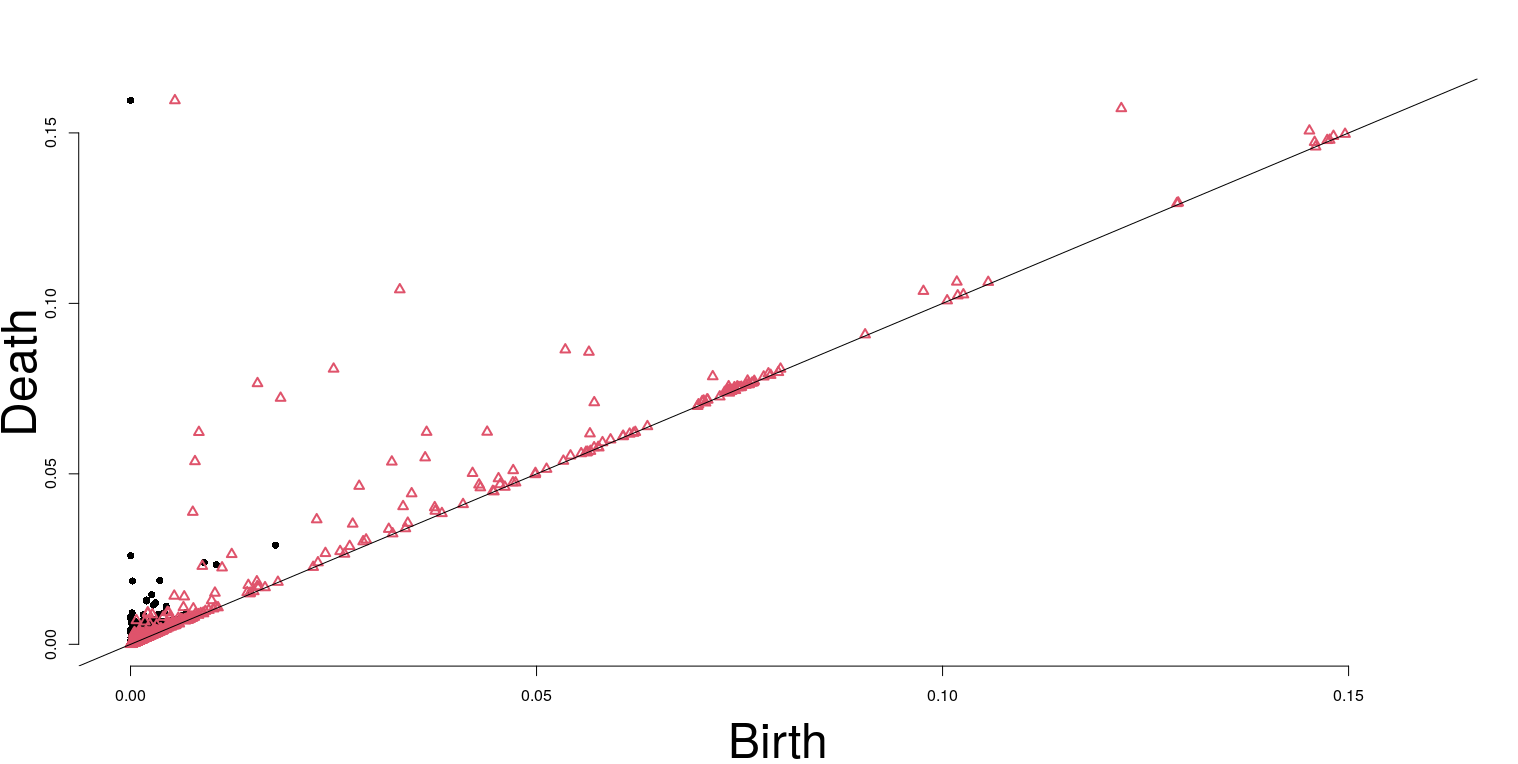}}
    \hfill
  \subfloat[Persistence diagram of the spectrogram's zeros\label{fig:diagram_zeros}]{\includegraphics[width=\linewidth]{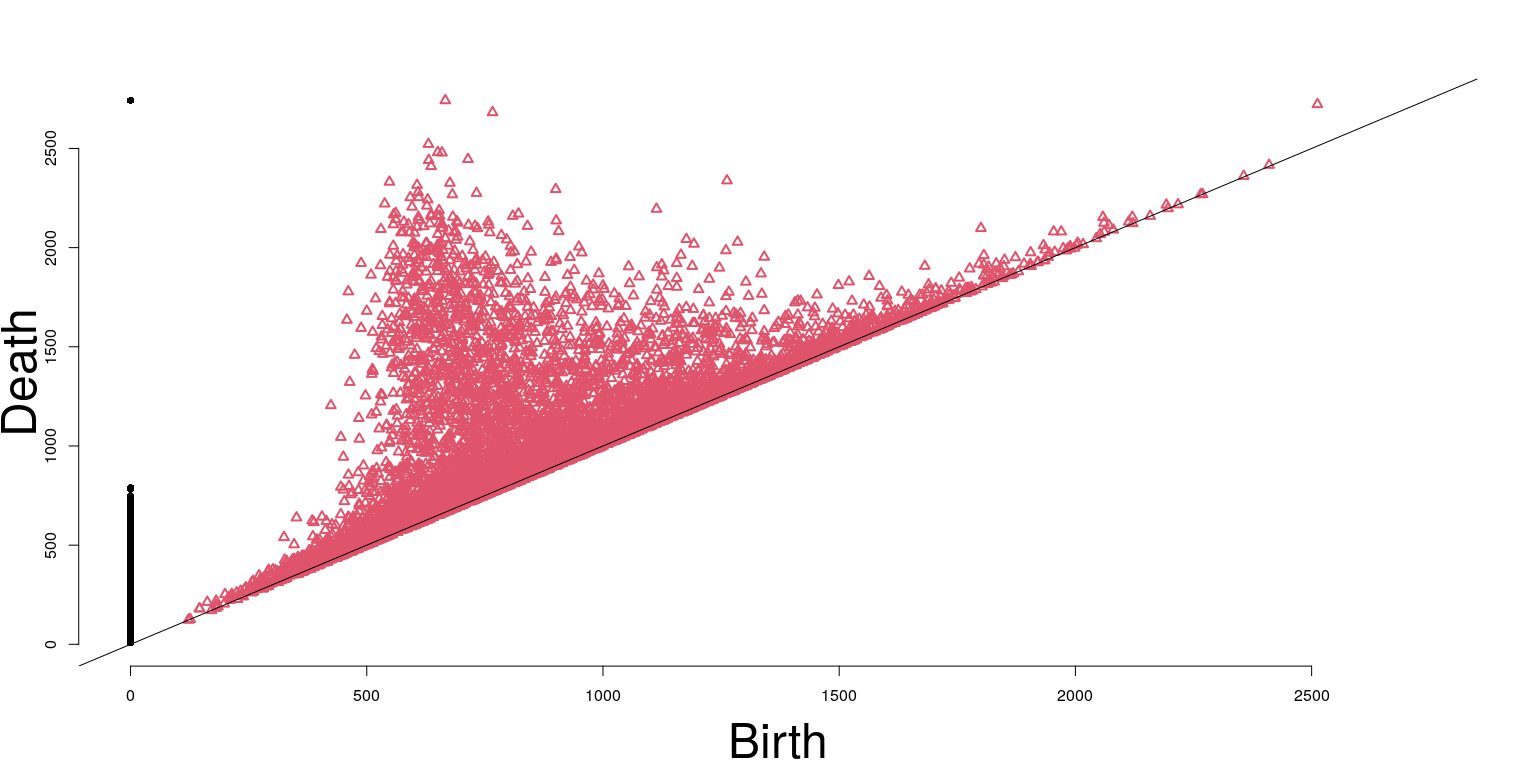}}
  \label{fig:diagrams}
  \hfill
    \subfloat[Persistence diagram of the Taken's embedding\label{fig:diagram_taken}]{\includegraphics[width=\linewidth]{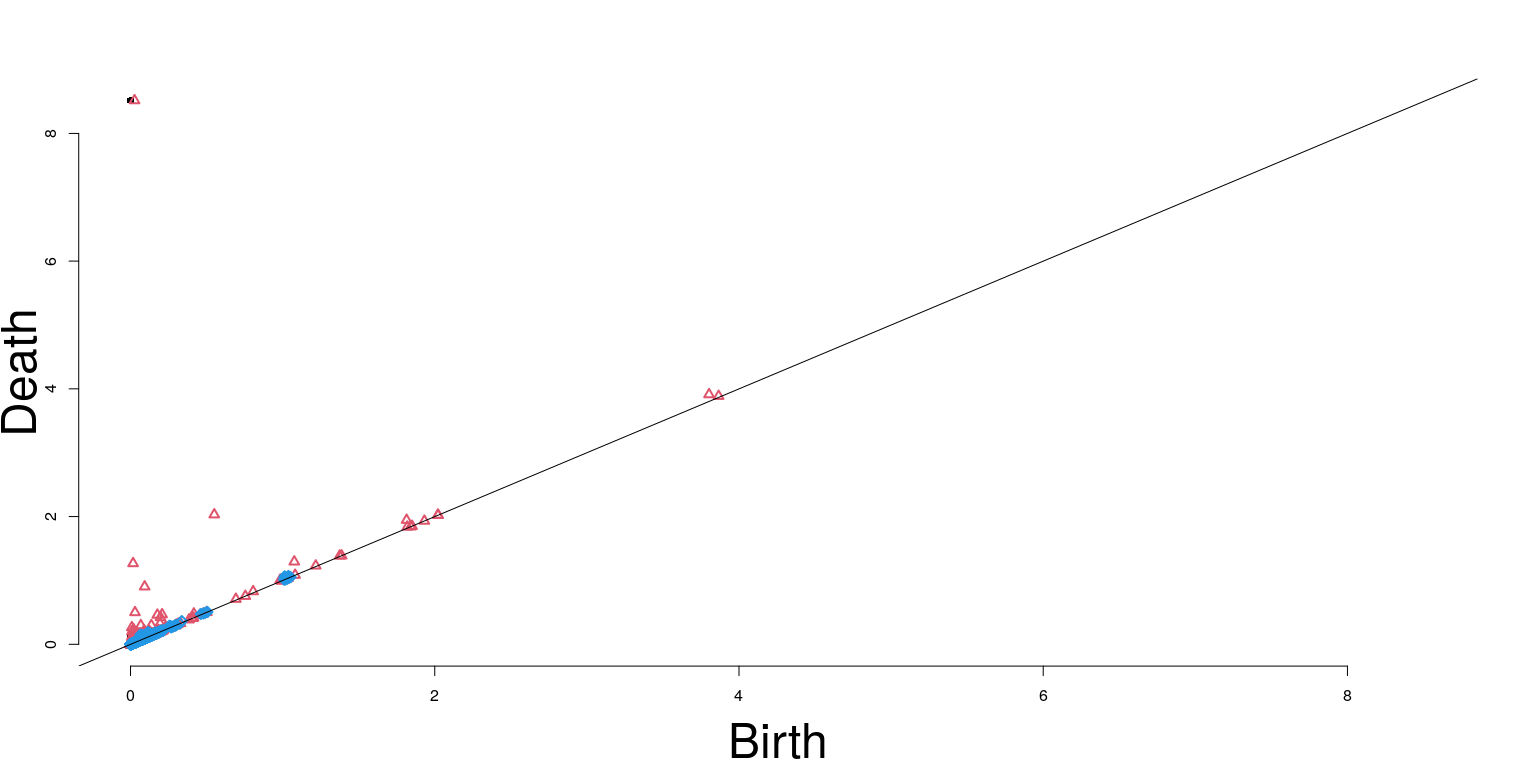}}
  \caption{Three persistent diagrams computed for the three different representation of the same signal of Figure \ref{fig:representation_signal}. \ref{fig:diagram_taken} is the persistence diagram of the spectrogram's surface using sublevel sets; \ref{fig:diagram_zeros} is the persistence diagram of the zeros of the spectrogram using an Alpha complex; \ref{fig:diagram_taken} is the persistence diagram of the Taken's embedding using an Alpha complex. The points in the diagram represent the computed persistent homologies. The different colors represent the different dimensions: black for $p=0$, red for $p=1$ and blue for $p=2$.}
\end{figure}

We described three ways of representing the same audio signal. It naturally raises some questions, such as, is there a representation that carries more topological (discriminant) information than the other? To get an intuition about this question, Figure~\ref{fig:diagrams} shows three persistence diagrams computed on the representations of the same signal illustrated on Figure~\ref{fig:representation_signal}  (see section~\ref{part:filtrations} for more information about how this topological information is computed). Although the signal is the same, the extracted topological characteristics are very different because of the choice of the representation.  Since the representation spaces can be of different dimensions, the homological features describing the topology of the object of these spaces will also be of different dimensions. This raises another, more specific questions, is the access to higher dimensions providing relevant information? We address these issues in a quantitative way, through a case study, a classification problem of human vowels.

\section{Topological Data Analysis: An overview}\label{part:theoretical_background}

This section presents the theory and data analysis pipeline of TDA in a nutshell. For more details, we refer the reader to two important textbooks that offer a fairly broad presentation of TDA and its theoretical foundations (i.e., \cite{edelsbrunnerComputationalTopologyIntroduction2009, deyComputationalTopologyData2022}) as well as the excellent introduction for data scientists \cite{chazalIntroductionTopologicalData2021}.

\subsection{Persistent homology}

In a prosaic summary, TDA is the enumeration of holes of different dimensions in the \emph{shape of data}. This intuitive notion of holes is formalized by the mathematical concept of a homology group. Homology groups allow us to treat the holes of a topological space mathematically, by studying the connectivity of the space. Let us denote as $H_p(\mathbb{M})$ the $p^{th}$ homology group of the topological space $\mathbb{M}$. Every homology group has the topological properties of its dimension. For $p=0$, $H_0(\mathbb{M})$ takes the connected elements of $\mathbb{M}$, for $p=1$, $H_1(\mathbb{M})$ takes the tunnels, for $p=2$, the voids.  The rank of $H_p(\mathbb{M})$ defines the $p^{th}$ Betti numbers, denotes as $\beta_p = rank(H_p)$ hereafter. Betti numbers basically count the number of different topological features. Homology groups and Betti numbers are topological invariants which characterizes the shape of $\mathbb{M}$. 

Instead of the homology groups of $\mathbb{M}$, we are interested in its persistent homology groups. We compute them through a filtration $F$ of $\mathbb{M}$. A filtration $F$ is a parametrized nested family of subspaces $F=(M_r)_{r\in T},$ where $T\subseteq\mathbb{R},$ such that for any $r,r'\in T,$ if $r\leq r',$ $M_r\subseteq M_{r'},$ and $\mathbb{M}=\cup_{r\in T} M_r$. The parameter $r\in T$ is the scale parameter. Let $f:\mathbb{M}\mapsto \mathbb{R}$  and $M_r=f^{-1}(-\infty,r]$ be the sublevel set for value $r$, the family $\left\{M_r\right\}_{r\in T}$ is the sublevel set filtration. A filtration allows multiscale topological description; for each value of $r$ we have the associated homology groups. The persistence of homologies makes it possible to track the lifetime of homologies, to determine their birth and their death according to the scale parameter $r$.

We denote as $\mathbb{X}$ our data objects that can be either $\mathcal{Z}_x$ or $\mathcal{P}_{x,D}$ (point clouds), or with the two-dimensional surface $\mathcal{S}_x^{(h)}$. A point cloud typically does not carry interesting topological information, nevertheless, it is possible to retrieve some via the construction of a simplicial filtration on the top of it. We can define a simplicial filtration by constructing a simplicial complex on the data. The filtration is then a nested sequence of simplicial complexes, for each of which we compute the (simplicial) homologies. In this paper, we use the Alpha complexes, a family of subcomplexes of the Delaunay complex, since it is equivalent but smaller than other complexes such as the Cech complexes. For any $x\in \mathbb{X}$, with $\mathbb{X}\subset\mathbb{R}^d$, and $r\in\mathbb{R}^+$, we define $B_x(r)=x+r \mathbb{B}^d$ the closed ball centered at $x$ and of radius $r$. The union of these balls is the set of points at a distance at most $r$ from at least one of the points of $\mathbb{X}$. We define the Alpha complex 
\begin{equation}\label{eq:alpha_complex}
    Alpha(r)=\{\sigma \subseteq \mathbb{X} | \cap_{x\in \sigma}R_x(r) \neq \emptyset\},
\end{equation}
where $R_x(r)$ is the intersection of each Euclidean ball with its corresponding Voronoi cell. In the context of filtration, we construct a nested sequence of Alpha complexes on $\mathbb{X}$, taking an increasing value of $r$.

We recover the homological information of $\mathbb{X}$ computed for different scale values $r$. The birth and death of the topological features (the elements that make up each homology group) are recorded and summarized in the persistence diagram, for each value of $r$. Thus, each persistent topological feature resulting from the filtration is expressed by a pair $(b_i,d_j)$, the moment of its birth and its death, i.e., two values of $r$. The most persistent homologies are those for which the difference $d_j-b_j$ is maximal \cite{fasyConfidenceSetsPersistence2014}. We define a persistence diagram as a multiset of points in $\mathcal{D}:= D\times\{1,...,P\}$, where 
\begin{equation}\label{eq:persistence_diagram}
    D:=\{(b,d)\in\mathbb{R}^2|d\geq b\geq 0\}.
\end{equation}
Each triplet $(b,d,p)\in\mathcal{D}$ represents a $p$-dimensional homological feature that appears when $r=b$ and disappears when $r=d$ \cite{maroulasBayesianFrameworkPersistent2020}.

In our experiment, we use filtration adapted to the representation spaces. More specifically, the alpha-complex filtration for Taken's embeddings and spectrogram zeroes and the sublevel set filtration for the spectrogram surface.

\subsection{Exploitation of the information from the diagram space}\label{part:vectorisation}

Persistence diagrams provide multiscale homological descriptions of the data. However, the space of persistence diagrams has very complicated geometry and topology which makes it difficult to exploit directly \cite{divol2021}. Many strategies were developed in order to extract the relevant information and enable machine learning applications. We refer the readers to \cite{barnesComparativeStudyMachine2021, henselSurveyTopologicalMachine2021} that offer comprehensive review of existing methods. 

Common approaches consist in computing functional representations alike persistent surfaces \cite{adamsPersistenceImagesStable2017}, persistence landscapes \cite{bubenikStatisticalTopologicalData2015} or persistent silhouettes \cite{chazalStochasticConvergencePersistence2014}, and then, to discretized them to be used as vector-based input for machine learning algorithms. These representations have been proven to enjoy stability properties \cite{bubenikPersistenceLandscapeIts2020a} meaning that they are slightly modified by small variations in the input data. Their nice properties and ease to compute make then widely employed in  applications. For example, \cite{liuApplyingTopologicalPersistence2016} uses  persistence landscapes as input to a CNN for musical signal classification, and  PLLay \cite{kimPLLayEfficientTopological2020} is a topological layer based on persistence landscapes.


Instead of functional representations, persistence diagrams can be summarized by scalar descriptors. There is a myriad of such, among which, the $p$-norm of the persistence diagram \cite{cohen-steinerLipschitzFunctionsHave2010} or the persistent entropy \cite{atienzaPersistentEntropySeparating2019}. Both are proven to be stable to perturbations of the input data \cite{atienzaStabilityPersistentEntropy2020}. Aggregating those descriptors together appears to be an interesting strategy.  \cite{carriereStableTopologicalSignatures2015} construct a stable vector of the persistence diagram from the distances between points and between each point and the diagonal, while  \cite{fireaizenAlarmSoundDetection2022} use a whole set of descriptors on the persistence diagrams. In the sequel, we will consider sets of descriptors as well as some standard functional summaries.

\section{Experiments}

In this section, we investigate how the choice of representation space of a digital audio signal affects topological information. This information is quantified in terms of Out Of Bag (OOB) error in a supervised classification problem inspired by the topical question in acoustic of vowel classification \cite{korkmazTurkishVowelClassification2019, georgiouComparisonPredictionAccuracy2023}. It finds applications in emotion classification \cite{debEmotionClassificationUsing2019}, in evaluating developmental trouble \cite{vavrinaDetectionDegreeDevelopmental2012} or in order to distinguish between healthy patients and patients with neurological disorder \cite{vashkevichClassificationALSPatients2021}. 

We collected our own dataset consisting of French-speaking adults pronouncing vowels indoors. For each of these recordings, we extract homological information from three possible representation spaces, as well as classical frequency descriptors from the associated spectrogram. We identify the recordings with the pronounced vowel, the individual pronouncing it, and their gender. Thus, we can quantify whether topological descriptors provide additional information to the more conventional frequency descriptors; whether there is a difference in the topological information depending on the representation space; whether there is a better way of vectorizing the persistence diagrams. In order to visualize the extracted information in a more qualitative way, we also learn the underlying manifold from the set of all extracted topological features of each representation and visualize how the signals are distributed over each manifold.

In what follows, we describe the recorded data and three associated classification problems on which the test is carried out. Then, we present the three representation spaces and associated filtration we use for each signal. Finally, we present different procedures for vectorization of persistent diagrams and the methodology we use to compare them.

\subsection{Presentation of the data}

The data are digital audio recordings of French-speaking adults. These recordings were made in a controlled environment to keep low and stable signal-to-noise ratio and to avoid junk sounds. We used a Zoom H6 recording device with a stereophonic microphone XYH-6. The sampling rate of the recordings is 44100 Hz, 16 bits. For the present analyses, we subsampled the signals to 16kHz and convert them to mononophonic. We recorded 20 individuals, 15 women and 5 men. Each individual pronounced 8 vowels \textipa{(ä, \~\textscripta, \textschwa, i, o, \~O, u, y)} in 7 conditions: natural, low voice, high voice, short, long, on an ascending and descending scale. There were 10 utterances for each condition. Therefore, each vowel has been recorded $1400$ times and each individual $560$ times. In total, $8400$ recordings were made by females and $2800$ by males. The audio data set is freely accessible \cite{bonafos_guillem_2023_7961904} and can be downloaded on \href{https://zenodo.org/record/7961904}{https://zenodo.org/record/7961904}.

\subsection{Comparison on three supervised classification problems}

We consider the three following problems : prediction of the vowel, prediction of the individual who pronounced the vowel, prediction of the gender of the individual who pronounced the vowel. For each problem (and each representation space), we use a random forest as a classification model with the same number of $500$ trees. We report the OOB, which allows us to estimate the error by cross-validation at a lower cost, for models using either topological variables only, or frequency variables only, or both topological and frequency variables (topologically-augmented).  

\subsection{Computation of the homological information}\label{part:filtrations}

For each record, we compute the spectrogram with a Gaussian window of $11.6$ ms and an overlap of $90\%$. On the spectrogram's surface representation, we apply sublevel set filtration to compute the persistent homologies while on the spectrogram's zeroes, we use an alpha-complex filtration (see section~\ref{part:theoretical_background}).

For the representation using Taken's embeddings, we first estimate for each record the two parameters required to build the embedding, $\tau$ according the AMI and $D$ according to the Cao's algorithm. Then, we calculate the embeddings of each record according to these two values. Finally, we harmonize the dimension of embedding spaces over recordings by reducing the dimension to 3 via UMAP \cite{mcinnesUMAPUniformManifold2020}. The records are represented by point clouds in $\mathbb{R}^3$. On this space, we compute the persistent homologies using an alpha-complex filtration.

At the end, each audio signal is associated to three persistent diagrams carrying potentially different topological information. 

We also compute the Mel-Frequency Cepstral Coefficients (MFCC) \cite{chachada_kuo_2014, sueurSoundAnalysisSynthesis2018}. Those are more classical frequency descriptors of the signal, specifically designed for human speech analysis and generally used in machine learning for classification tasks involving human speech. They will serve as a baseline to compare our topological descriptors. Finally, in a topologically-augmented machine learning fashion, we will merge the MFCC with the topological descriptors.

\subsection{Extraction of the information from the persistence diagram and comparison of the persistent variables}

As discussed in part~\ref{part:vectorisation}, there are many ways to summarize the information carried by persistence diagrams into variables that facilitate further statistical/machine learning usage. We do not claim to be exhaustive in our comparison. We follow various proposals from the literature to form a set of variables computed on persistence diagrams, here so-called \emph{persistent variables}. In addition to persistent variables, we also compute functional summaries. To be fair in comparison, we use the same classification model in each case.

\subsubsection{Persistent variables}

Let $\mathcal{D}$ a persistence diagram, the multiset defined in equation~(\ref{eq:persistence_diagram}). We compute each variable for $p=\left\{0, 1\right\}$, for the Taken's embeddings we also compute for $p=2$.

From \cite{atienzaPersistentEntropySeparating2019}, we compute the persistent entropy $E_p$. Let $L_p= \{\ell_i = d_i - b_i | 1 \leq i \leq n \}$ the set containing the lifetime of each homological features of dimension $p$ of the diagram $D$. We define the persistent entropy as 
\begin{equation}\label{eq:persistent_entropy}
    E_p(F) = -\sum\limits_{i=1}^n p_i\log (p_i),    
\end{equation}
where $p_i = \frac{\ell_i}{S_L}$ and $S_L = \sum\limits_{i=1}^n \ell_i.$ We compute it for each dimension.

We follow \cite{fireaizenAlarmSoundDetection2022} to compute several descriptors of the persistence diagram. Using $L_p$, we compute some statistics from the vector:
\begin{enumerate}[label=\roman*)]
    \item\label{eq:mu} the mean $\mu_p$.
    \item\label{eq:sigma2} the variance $\sigma_p^2$.
    \item\label{eq:5_longest_lifetime} the 5 top longest lifetimes $L_{p,i}$ for $i=1,...,5$.
    \item\label{eq:normalized_longest_1} two normalized longest lifetime, 
    $$\frac{L_{p,1}}{|L_p|}$$.
    \item\label{eq:normalized_longest_2} and
    $$\frac{L_{p,1}}{\mu_p}.$$
    \item\label{eq:n_alpha} the number of $\alpha$-long lifetime $N_{p, \alpha}$. We choose $\alpha$ to distinguish between topological noise and information from the diagram. We take $\alpha=0.05$. 
    $$N_{p, \alpha} = \#(L_{p,i} > \alpha).$$
    \item\label{eq:ratio_mu} the ratio of means 
    $$\frac{\mu_{p_1}}{\mu_{p_2}}.$$ 
    $p_1$ and $p_2$ being two dimensions of homological features, with $p_1 < p_2$.
    \item\label{eq:ratio_n_alpha} the ratio of $\alpha$-long cycles 
    $$\frac{N_{p_1, \alpha}}{N_{p_2, \alpha}}.$$
    \item\label{eq:product_longest} the products of top longest lifetimes 
    $$L_{p_1, i} \cdot L_{p_2,i},$$ for $i=1,...,6$.
    \item\label{eq:product_difference_longest} the products 
    $$L_{p_1,i} \cdot  (L_{p_2,i} - L_{p_2,i+1})$$ for $i=1,...,6$.
    \item\label{eq:periodicity_score} the Periodicity Score 
    $$PS = 1-\frac{L_{1,2}}{L_{1,1}}.$$
    \item\label{eq:quasiperiodicity_score} the Quasi-Periodicity Score 
    $$QPS = L_{1,2} \cdot L_{2,1}.$$
    \item\label{eq:frequency_score} the Frequency Shift Score 
    $$FSS = \frac{L_{2,1}\cdot L_{2,2}}{L_{1,1}}.$$
\end{enumerate}

We compute four more variables from \cite{pereiraPersistentHomologyTime2015}: 
\begin{itemize}
    \item the persistent Betti number. For any pair of indices $0\leq k\leq l\leq n$ and any dimension $p$, the $p^{th}$ persistent Betti number is 
    \begin{equation}\label{eq:persistent_betti}
        \beta_p^{k,l}=\sum_{i\leq k}\sum_{j>l}\kappa_p^{i,j},
    \end{equation}
    where $\kappa_p^{i,j}$ is the number of $p$-dimensional homology that are born at $X_i$ and die at $X_j$, $X_i$ and $X_j$ being two subsets of the filtration. The persistent Betti number is the number of holes for each dimension, and is called so in \cite{pereiraPersistentHomologyTime2015}. We follow \cite{edelsbrunnerComputationalTopologyIntroduction2009} and call it the persistent Betti number ;
    \item the maximum hole lifetime in each dimension 
    \begin{equation}\label{eq:max_lifetime_hole}
        \text{max}_p = \max_{\ell_i\in L_p}(\ell_i);
    \end{equation}
    \item the number of relevant holes 
    \begin{equation}\label{eq:n_relevant}
        \text{n\_rel}_p = \sum_{\ell_i\in L_p}f(\ell_i, \max_p, \text{ratio}),
    \end{equation} 
    where $f(\ell_i, \max_p, \text{ratio})$ equals 1 if $\ell_i \geq \max_d \times \text{ratio},$ 0 otherwise. It is the number of points in the persistent diagram that are relatively distant from the diagonal. We chose to count the holes with a lifetime at least greater than a quarter of the longest (\textit{i.e.}, $\text{ratio} = 0.25$) ;
    \item the sum of all lifetime 
    \begin{equation}\label{eq:sum_lifetime}
        \text{sum}_p= \sum_{\ell_i\in L_p}(\ell_i).
    \end{equation}
\end{itemize}

Following \cite{cohen-steinerLipschitzFunctionsHave2010}, we compute the $p$-norm of the persistence diagram, 
\begin{equation}\label{eq:norm_persistence_digram}
    \lVert D \rVert_p = [\sum\limits_{u\in D} \text{pers}(u)^p]^{\frac{1}{p}},    
\end{equation}
where $u$ is a point of the diagram and $\text{pers}(u)$ the absolute value of the difference between the coordinates. In practice, we compute it for $p=2$.

The set of persistent variables is used alone in the different classification problems, as well as in combination with the MFCCs. This data fusion method, which we expect to be the most efficient according to the results from the literature, can be found under the name "Topology augmented" in Table~\ref{tab:oob}, which summarizes the results for the different problems.

\subsubsection{Functional summary of the persistence diagrams}

In addition to the set of persistent variables, we consider functional summary of the persistent diagrams. After being discretized, they can use it as input of a classification model. We present and test two of these methods.

First, we compute the silhouette of a persistence diagram \cite{chazalStochasticConvergencePersistence2014}. It follows the persistence landscapes \cite{bubenikStatisticalTopologicalData2015, bubenikPersistenceLandscapeIts2020a}. The silhouette summarizes the persistence diagram in a single function. We choose this representation to illustrate the mapping of the persistence diagram into a Hilbert space. We define it as 
\begin{equation}\label{eq:silhouette}
    \phi^{(\gamma)}(t)=\frac{\sum_{j=1}^m|d_j - b_j|^\gamma \lambda_j(t)}{\sum_{j=1}^m|d_j - b_j|^\gamma}.
\end{equation} 
The silhouette takes a parameter $\gamma$. This determines whether all points are treated equally ($\gamma$ small) or whether the most persistent pairs of points are given more weight ($\gamma$ large). We set $\gamma=1$, because some results show the importance of what is sometimes considered as "topological noise" (e.g., \cite{patrangenaruChallengesTopologicalObject2019}). We also have to fix the number of sample on which we build the silhouette. We set $\text{nsample}=2^9=512$.

Next, we compute the persistence images of the diagrams \cite{adamsPersistenceImagesStable2017}. The idea is to map the persistence diagram to an integrable surface, so-called the persistence surface 
$$\rho_D (u,v) = \sum_{(x,y)=(b, d-b)\in D}w(x,y)g_{(x,y)}(u,v),$$ 
where $g_{(x,y)}(u,v):\mathbb R^2 \rightarrow \mathbb R$ is bivariate Gaussian distribution centered at each point $(x,y)=(b, d-b) \in D$ and $w:\mathbb{R}^2\mapsto\mathbb{R}$ is a continuous and piecewise differentiable weighting function. $(u,v) \in \mathcal D$ is a compact domain, \textit{e.g.}, the domain of definition of the points $(x,y)=(b,d-b) \in D$. Then, we divide the domain in a collection of non-overlapping subdomains, the pixels $P_i$, with $\mathcal D = \bigcup P_i$. We integrate the persistence surface over the fixed grid to define the persistence image, taking the average of $\rho$ in each pixel, 
\begin{equation}\label{eq:persistence_image}
    I_{P_i}(\rho_D) = \int\int_{P_i}\rho_D (u,v) du dv.    
\end{equation}
This outputs an image representing the persistence diagram, with a density distribution that is more or less important depending on the distribution of homologies on the persistence diagram. Again, we can use these vectors as input to a classification model. A sampling parameter must be set, which is fixed at 10.

\subsection{Step-wise selection of the variables}\label{part:stepwise}

As aforementioned in Section~\ref{part:vectorisation}, the question of how to extract information from the persistence diagram for statistical purposes remains widely open and is not to being addressed in this paper. We instead focus on the study of the differences in persistent homologies between representation spaces of a given signal, and we bring  complementary information to the field by comparing the most frequently occurring topological variables.

In order to carry out this comparison, we follow a step-wise strategy. For each classification problem, we start with the complete set of persistent variables. We estimate the model with this set of variables, and then we remove the least important variable. We retrain the model with this new set of variables. This process continues until we remove all the variables in the training set. In the end, we retrain the model with the smallest OOB error. We report in Table~\ref{tab:oob} the results of the best model, and we list the variables present in the set of variables used to train this model. We count each time a variable is in the set of the best model, for each problem and for each representation.

This information can be found in the Table~\ref{tab:persistent_variables}. We can then compare whether there is a difference according to the input representation, whether certain topological variables carry more interesting information than others, or at least whether they are more often found in the best model for the classification tasks in question. Since our procedure includes MFCC, we potentially remove some MFCC descriptors while keeping some topological variables in models producing the best results. 

There are three classification problems (prediction of the vowel, of the gender, and of the individual), three initial signal representations (spectrogram's surface, spectrogram's zeros, taken's embeddings). For each, we follow our step-wise strategy to find the best model, eliminating iteratively the less useful variable, with two conditions: with and without the MFCC. We save the $18$ best models, and we count the remaining persistent variables in each of them. Lastly, we follow the same procedure by including the topological variables of all representations in the same model, for the three problems. Thus, a variable can be counted in a maximum of 24 models.

We put in bold in Table~\ref{tab:persistent_variables} the variables that appear at least in 50\% of the best models. In the signal representation columns, each variable can appear at best 12 times. Here again, we put in bold those that appear at least in 50\% of the best models. This way, it will be possible to compare whether certain variables appear more than others globally, whatever the representation and whatever the problem. We can also check if there are differences depending on the representation space. For the sake of clarity, we resume the top 5 longest lifetimes $L_{p,i}$ and their product $L_{p_1,i}\cdot L_{p_2,i+1}$ by a row in  Table \ref{tab:persistent_variables} for each dimension. Also, the maximum count for these variables is not 24, but $24\times 5=120$.

\section{Results}

\subsection{Supervised problem}

\begin{table}[ht]
\begin{threeparttable}
\caption{Comparison of Out Of Bag error (OOB) for different signal representations}
\label{tab:oob}
\setlength\tabcolsep{0pt} 

\begin{tabular*}{\columnwidth}{@{\extracolsep{\fill}} ll cccc}
\toprule
     Signal Representation & Vectorization\tnote{a} & \multicolumn{3}{c}{OOB (\%)} \\ 
\cmidrule{3-5}
     & & Vowel & Gender & Individual \\
\midrule
     MFCC & & 8.71 & \textbf{4.54} & 11.54 \\
\addlinespace
     Spectrogram's Surface & Silhouettes $p = 0$ & 72.97 & 20.37 & 73.72 \\
     & Silhouettes $p=1$ & 46.91 & 19.34 & 63.34\\
     & Silhouettes $p=0,1$ & 45.04 & 16.44 & 53.86\\
     & Persistent Image $p=0$ & 79.31 & 28.46 & 88.27\\
     & Persistent Image $p=1$ & 79.26 & 28.73 & 87.97\\
     & Persistent Image $p=0,1$ & 76.99 & 26.1 & 83.43\\
     & Persistent Variables & 52.03 & 15.72 & 50.07\\
     & Topology augmented\tnote{b} & 8.43 & 4.71 & 10.42\\
\addlinespace
     Spectrogram's Zeros & Silhouettes $p=0$ & 79.89 & 23.69 & 78.21\\
     & Silhouettes $p=1$ & 73.33 & 17.61 & 75.69\\
     & Silhouettes $p=0,1$ & 70.11 & 16.14 & 67.5\\
     & Persistent Image $p=0$ & 82.48 & 25.08 & 87.77\\
     & Persistent Image $p=1$ & 70.5 & 16.79 & 71.15\\
     & Persistent Image $p=0,1$ & 70.7 & 17.19 & 70.68\\
     & Persistent Variables & 69.85 & 15.19 & 62.15\\
     & Topology augmented\tnote{b} & 8.03 & 5.55 & \textbf{9.24}\\
\addlinespace
     Taken's embeddings & Silhouettes $p=0$ & 84.39 & 27.71 & 89.15\\
     & Silhouettes $p=1$ & 81.51 & 25.32 & 84.44\\
     & Silhouettes $p=2$ & 79.71 & 25.02 & 81.19\\
     & Silhouettes $p=0,1$ & 78.03 & 24.59 & 80.63\\
     & Silhouettes $p=0,2$ & 78.79 & 24.56 & 77.86\\
     & Silhouettes $p=1,2$ & 78.56 & 24.68 & 77.22\\
     & Silhouettes $p=0,1,2$ & 76.41 & 23.98 & 75.53\\
     & Persistent Image $p=0$ & 85.11 & 29.63 & 90.24\\
     & Persistent Image $p=1$ & 81.89 & 27.06 & 84.87\\
     & Persistent Image $p=2$ & 83.5 & 27.15 & 87.42\\
     & Persistent Image $p=0,1$ & 80.76 & 26.29 & 82.42\\
     & Persistent Image $p=0,2$ & 81.54 & 25.94 & 82.93\\
     & Persistent Image $p=1,2$ & 80.26 & 26.21 & 82.24\\
     & Persistent Image $p=0,1,2$ & 79.53 & 25.76 & 79.86\\
     & Persistent Variables & 69.85 & 20.27 & 60.89\\
     & Topology augmented\tnote{b} & 8.49 & 4.79 & 10.77\\
\addlinespace
      All together & Persistent Variables & 41.56 & 10.89 & 32.13 \\
      & Topology augmented\tnote{b} & \textbf{7.98} & 6.01 & 10.3\\
\bottomrule
\end{tabular*}

\smallskip
\scriptsize
\begin{tablenotes}
\RaggedRight
\item[a] Strategy to extract information from the persistence diagram
\item[b] Signal representation = MFCC + persistent variables. We follow the step-wise strategy presented in section~\ref{part:stepwise} on this set and present the result of the best model.
\end{tablenotes}
\end{threeparttable}
\end{table}

The results for all supervised classification problems can be found in Table~\ref{tab:oob}. All models are random forests with the same number of trees. For models using as covariates either persistent variables or both persistent variables and MFCCs, the reported results are those of the best model (i.e., the model following the stepwise procedure, enjoying the lowest OOB).

\subsubsection{TDA is useful for signal classification}

Topological information improves the results of two over three classification problems. For vowel classification, the MFCCs alone obtain an OOB of 8.71\%. The addition of topological information improves the results, whatever the chosen representation space. The OOB reduces to 8.43\%, 8.03\%, 8.49\% using the persistent variables obtained from spectrogram's surfaces, spectrogram's zeros and Taken's embeddings, respectively. The best improvement is obtained by taking all persistent variables from all representation spaces, resulting in an OOB of 7.98\%.

The individuals' classification, also benefits from topological information. Indeed, MFCCs alone are outperformed when adding topological information, whatever the representation space. The best results are obtained when the persistent variables are extracted from the spectrogram's zeros, lowering the OOB from 11.54\%  to 9.24\%. On the contrary to vowel classification, taking all persistent variables from all representation spaces deteriorates the results.

Finally, on the gender classification problem, topology augmented approach fails to improve the results over the MFCCs alone, which exhibits the lowest OOB at 4.54\%.  

\subsubsection{Topological information alone is not enough to discriminate the signal}


Topological information alone, whatever the chosen representation space, is outperformed by classical frequency descriptors on each problem. For the vowel classification problem, while the MFCCs alone achieve an OOB of 8.71\%, the best result using only topological variables is of about 41\%. It is obtained with persistent variables from all representations. 

For the gender classification problem, using MFCCs alone give the best result, with an OOB of 4.54\%. Using topological information, the best models always consider the persistent variables, with OOB ranging from 15.19\% to 20.27\%. Aggregating the persistent variables of each representation improves the results, with an OOB of 10.89\%.

For the individual classification problem, it is once again the set of persistence variables of all three representations, that obtains the best results among models considering only topological information. It performs poorly with an OOB of 32.13\% while MFCCs alone reach OOB of 11.54\%.


\subsubsection{Complementarity of representations}

Topology augmented approaches improve the results over the MFCCs for both vowel and individual classification. Comparing the topology augmented approaches for the three representations on the vowel classification problem, it is the variables extracted from the persistence diagrams of the spectrogram's zeros that achieve the minimum OOB. For the individual classification problem, the variables extracted from the persistence diagrams of the zeros of the spectrograms also provide the best score for all the approaches compared for this problem. 


While the topology augmented approach with the addition of the persistent variables of the spectrogram's zeros provides the best improvement for both problems, there does not appear to be one better representation of the signal or one with a more informative topology than the others. The differences are held and, for the gender classification problem, the topology augmented models with the persistence diagrams of the spectrogram's surfaces and Taken's embeddings outperform the spectrogram's zeros. For the vowel classification problem, the best model is the one taking the full set of persistent variables in addition to the MFCCs, those computed on all representations.




The resulting 'data objects' have different topology, which would be potentially complementary. Indeed, we note that, for each problem, the model learned with the set of all persistent variables, computed on the three representations, and without the MFCCs, performs better than each model learned with the set of persistent variables computed on each representation. Thus, for vowel classification, while the persistent variables of the spectrogram's surfaces have an OOB of 52.03\%, those of the spectrogram's zeros 69.85\% and those of the Taken's embeddings 69.85\%, the model trained on the set of these persistent variables reaches an OOB of 41.56\%. Moreover, by adding the MFCCs, it is this set that gives the best results for this problem. For gender classification, the persistent variables computed on the spectrogram's surfaces have an OOB of 15.72\%, those on the spectrogram zeros 15.19\% and those on the Taken's embeddings 20.27\%. All together, the model still obtains a clear improvement, with an OOB of 10.89\%. For classification of the individuals, the persistent variables computed on the surfaces of the spectrogram have an OOB of 50.07\%, those of the zeros of the spectrogram 62.15\%, those of the Taken's embdedings 60.89\%. The model learned on the whole of these variables obtains an OOB of 32.13\%. Thus, we improve each time the results by merging with all the persistent variables.

\subsection{Best topological variables}

\begin{table*}
\begin{threeparttable}
\caption{Most frequently kept persistent variables in each best model}
\label{tab:persistent_variables}

\begin{tabularx}{\textwidth}{XX *{3}X}
\toprule
     Variable & Frequency in the best model (\%)\tnote{a} & \multicolumn{3}{c}{Signal Representation} \\ 
\cmidrule{3-5}
     & & Spectrogram's Surface & Spectrogram's Zeros & Taken's embeddings \\
\midrule
     $\beta_0$ \qquad \eqref{eq:persistent_betti} & \textbf{62.5} & 33.33 & 41.67 & \textbf{50} \\
     $\beta_1$ & 33.33 & \textbf{50} & 0 & 16.67 \\
     $\beta_2$ & 25 & $\emptyset$ & $\emptyset$ & \textbf{50} \\
     $\max_0$ \qquad \eqref{eq:max_lifetime_hole} & 29.17 & \textbf{58.33} & 0 & 0 \\
     $\max_1$ & 0 & 0 & 0 & 0 \\
     $\max_2$ & 0 & $\emptyset$ & $\emptyset$ & 0\\
     $\text{n\_rel}_0$ \qquad \eqref{eq:n_relevant} & 0 & 0 & 0 & 0 \\
     $\text{n\_rel}_1$ & 4.17 & 0 & 1 & 0 \\
     $\text{n\_rel}_2$ & 0 & $\emptyset$ & $\emptyset$ & 0 \\
     $\text{sum}_0$ \qquad \eqref{eq:sum_lifetime} & 41.67 & \textbf{66.67} & 0 & 2 \\
     $\text{sum}_1$ & \textbf{50} & 5 & 2 & 5 \\
     $\text{sum}_2$ & 8.33 & $\emptyset$ & $\emptyset$ & 16.67 \\
     $E_0$ \qquad \eqref{eq:persistent_entropy} & 29.17 & \textbf{50} & 0 & 8.33 \\
     $E_1$ & 33.33 & \textbf{50} & 0 & 16.67 \\
     $E_2$ & 4.17 & $\emptyset$ & $\emptyset$ & 8.33 \\
     $\lVert D_0 \rVert_p$ \qquad \eqref{eq:norm_persistence_digram} & 37.5 & \textbf{50} & 16.67 & 8.33 \\
     $\lVert D_1 \rVert_p$ & \textbf{50} & 25 & \textbf{50} & 25 \\
     $\lVert D_2 \rVert_p$ & 0 & $\emptyset$ & $\emptyset$ & 0 \\
     $\mu_0$ \qquad (\ref{eq:mu} & \textbf{87.5} & \textbf{83.33} & \textbf{75} & 16.67\\
     $\mu_1$ & \textbf{79.17} & \textbf{66.67} & \textbf{50} & 41.67\\
     $\mu_2$ & 4.17 & $\emptyset$ & $\emptyset$ & 8.33 \\
     $\sigma^2_0$ \qquad (\ref{eq:sigma2} & 45.83 & 33.33 & 41.67 & 16.67 \\
     $\sigma^2_1$ & \textbf{66.67} & 25 & \textbf{100} & 8.33 \\
     $\sigma^2_2$ & 0 & $\emptyset$ & $\emptyset$ & 0 \\
     $\frac{L_{0,1}}{|L_0|}$ \qquad (\ref{eq:normalized_longest_1} & 25 & 33.33 & 0 & 16.67 \\
     $\frac{L_{1,1}}{|L_1|}$ & 33.33 & \textbf{58.33} & 0 & 8.33 \\
     $\frac{L_{2,1}}{|L_2|}$ & 0 & $\emptyset$ & $\emptyset$ & 0 \\
     $\frac{L_{0,1}}{\mu_0}$ \qquad \ref{eq:normalized_longest_2} & \textbf{54.17} & 41.67 & 41.67 & 25 \\
     $\frac{L_{1,1}}{\mu_1}$ & 37.5 & \textbf{50} & 8.33 & 16.67 \\
     $\frac{L_{2,1}}{\mu_2}$ & 0 & $\emptyset$ & $\emptyset$ & 0 \\
     $N_{0, \alpha}$ \qquad (\ref{eq:n_alpha} & 29.17 & 0 & \textbf{58.33} & 0 \\
     $N_{1, \alpha}$ & 25 & 0 & 8.33 & 41.67 \\
     $N_{2, \alpha}$ & 0 & $\emptyset$ & $\emptyset$ & 0 \\
     $\frac{\mu_0}{\mu_1}$ \qquad (\ref{eq:ratio_mu} & \textbf{87.5} & \textbf{100} & 33.33 & 41.67 \\
     $\frac{\mu_0}{\mu_2}$ & 12.5 & $\emptyset$ & $\emptyset$ & 25 \\
     $\frac{\mu_1}{\mu_2}$ & 4.17 & $\emptyset$ & $\emptyset$ & 8.33 \\
     $\frac{N_{0, \alpha}}{N_{1, \alpha}}$ \qquad \ref{eq:ratio_n_alpha} & 20.84 & 0 & 33.33 & 8.33\\
     $\frac{N_{0, \alpha}}{N_{2, \alpha}}$ & 0 & $\emptyset$ & $\emptyset$ & 0 \\
     $\frac{N_{1, \alpha}}{N_{2, \alpha}}$ & 4.17 & $\emptyset$ & $\emptyset$ & 8.33 \\
     $PS$ \qquad (\ref{eq:periodicity_score} & 12.5 & 25 & 0 & 0 \\
     $QPS$ \qquad (\ref{eq:quasiperiodicity_score} & 0 & $\emptyset$ & $\emptyset$ & 0\\
     $FSS$ \qquad (\ref{eq:frequency_score} & 0 & $\emptyset$ & $\emptyset$ & 0\\
\addlinespace
     $L_{0,i}\tnote{b}$ \qquad (\ref{eq:5_longest_lifetime} & 21.67 & 36.67 & 5 & 1.67 \\
     $L_{1,i}\tnote{b}$ & \textbf{50.83} & 43.33 & \textbf{51.67} & 6.67 \\
     $L_{2,i}\tnote{b}$ & 5 & $\emptyset$ & $\emptyset$ & 10 \\
     $L_{0, i} \cdot L_{1, i}\tnote{b}$ \qquad (\ref{eq:product_longest} & 27.5 & 31.67 & 15 & 8.33 \\
     $L_{0, i} \cdot L_{2, i}\tnote{b}$ & 0 & $\emptyset$ & $\emptyset$ & 0\\
     $L_{1, i} \cdot L_{2, i}\tnote{b}$ & 2.5 & $\emptyset$ & $\emptyset$ & 5 \\
     $L_{0,i} \cdot  (L_{1,i} - L_{0,i+1})\tnote{b}$ \qquad (\ref{eq:product_difference_longest} & 5 & 5 & 3.33 & 1.67 \\
     $L_{0,i} \cdot  (L_{2,i} - L_{0,i+1})\tnote{b}$ & 0.3 & $\emptyset$ & $\emptyset$ & 1.67 \\
     $L_{1,i} \cdot  (L_{2,i} - L_{1,i+1})\tnote{b}$ & 0  & $\emptyset$ & $\emptyset$ & 0 \\

\bottomrule
\end{tabularx}

\smallskip
\scriptsize
\begin{tablenotes}
\RaggedRight
\item[a] We follow the step-wise strategy to choose the best model for each classification problem. we count in how many models each variable appears. 
\item[b] We count if one of them is kept, $\forall i \in \{1,...,6\}$
\end{tablenotes}
\end{threeparttable}
\end{table*}

For this classification problems, persistent silhouettes or persistent images perform worse than the persistent variables in all scenarios. For this reason, we build the topologically augmented approach only with the persistent variables. They are therefore not taken into account in the stepwise procedure to identify the best topological variable (see Section~\ref{part:stepwise}).. All the results are presented in Table~\ref{tab:persistent_variables}.


In general, there are 8 persistent variables that we encounter at least in 50\% of the best models. These are the persistent Betti number in dimension $p=0,$ the sum of $L_p$ in dimension $p=1,$ the 2-norm of the persistence diagram for $p=1,$ the average of $L_p$ for $p=0$ and $p=1$ and its variance for $p=1$, the normalized longest lifetime for $p=0,$ with the normalization computed with the mean of $L_p$ of this dimension, the ratio of the mean of $L_0$ and $L_1$.

The number of variables retained in 50\% of the best models varies significantly from one representation space to another. We identified $11,\ 5,\ 2$ persistent variables retained for the spectrogram's surface, for the spectrogram's zeros and for the Taken's embedding, respectively. 

There are two persistent variables that seem to stand out the most: the persistent Betti number and the norm of the persistence diagram. The other persistent variables which appear regularly are linked and summarized in the vector $L_p$. We find statistics on this vector that summarize the information of the persistence diagrams (mean, variance, normalized maximum), whatever the initial representation of the signal.

\subsection{Unsupervised analysis, learning the manifold}

\begin{figure*}[!t]
\centering
  \subfloat[MFCC Vowel\label{fig:umap_mfcc_vowel}]{\includegraphics[width=0.3\linewidth]{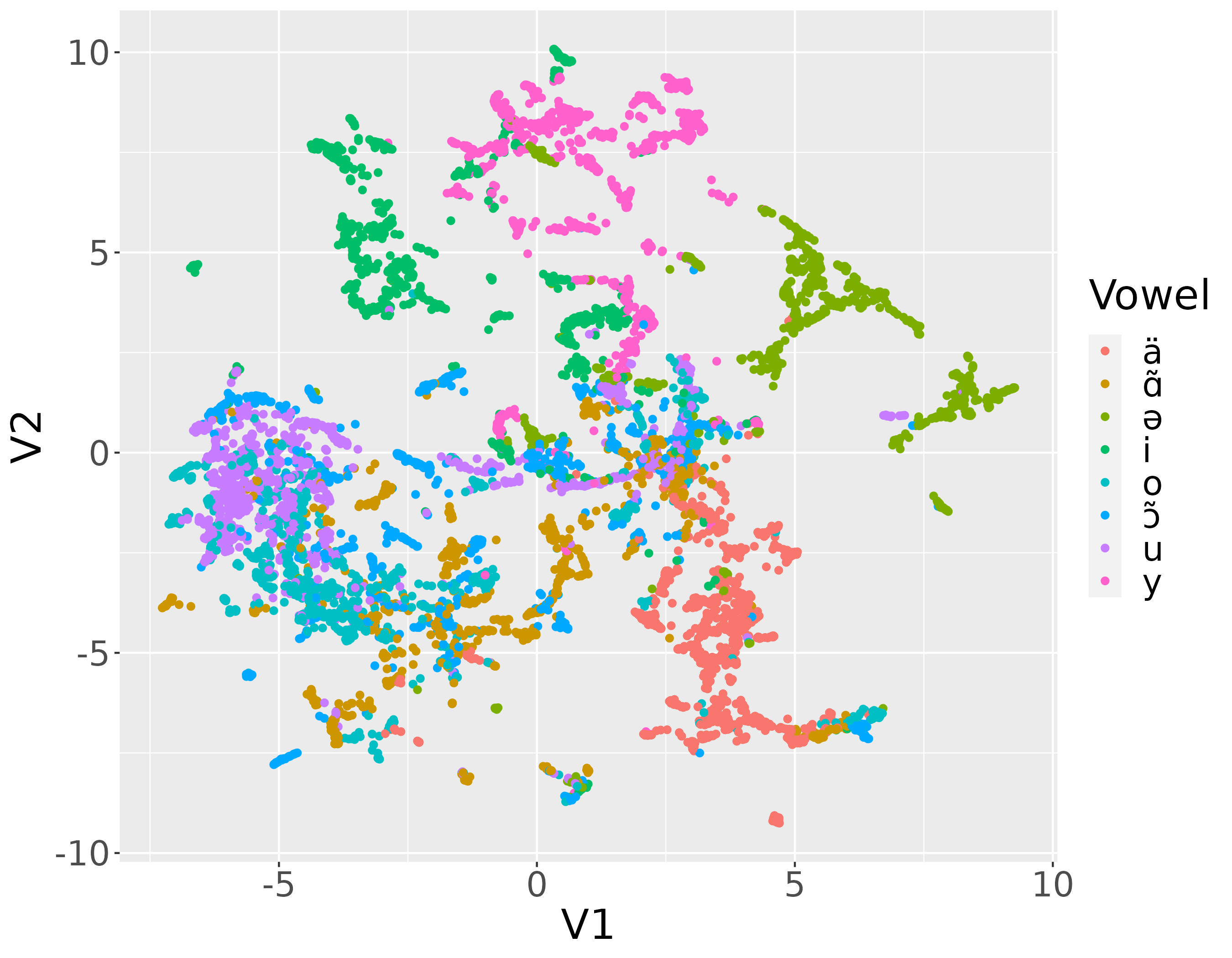}}
  \hfill
  \subfloat[MFCC Sex\label{fig:umap_mfcc_sex}]{\includegraphics[width=0.3\linewidth]{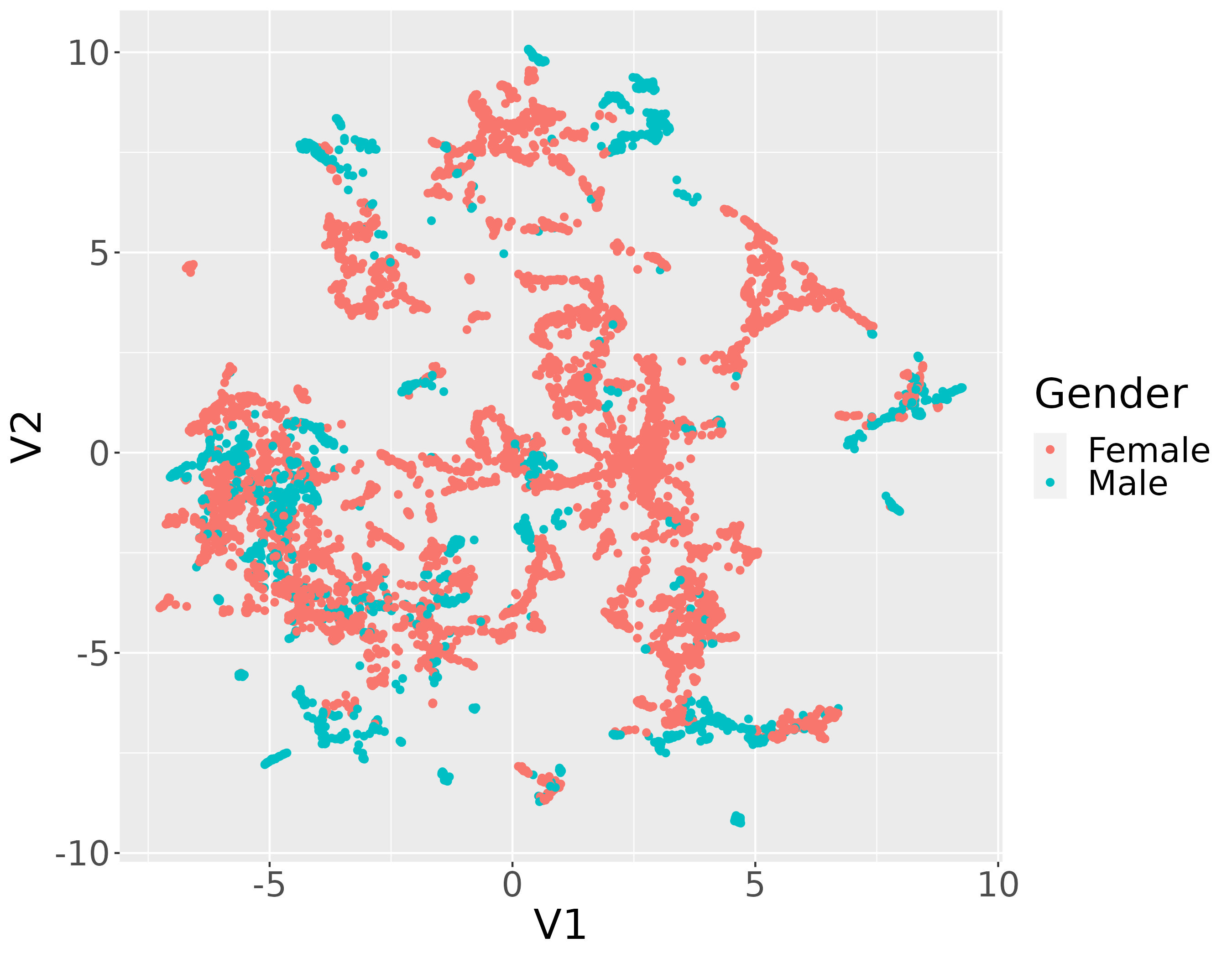}}
  \hfill
  \subfloat[MFCC subject\label{fig:umap_mfcc_subject}]{\includegraphics[width=0.3\linewidth]{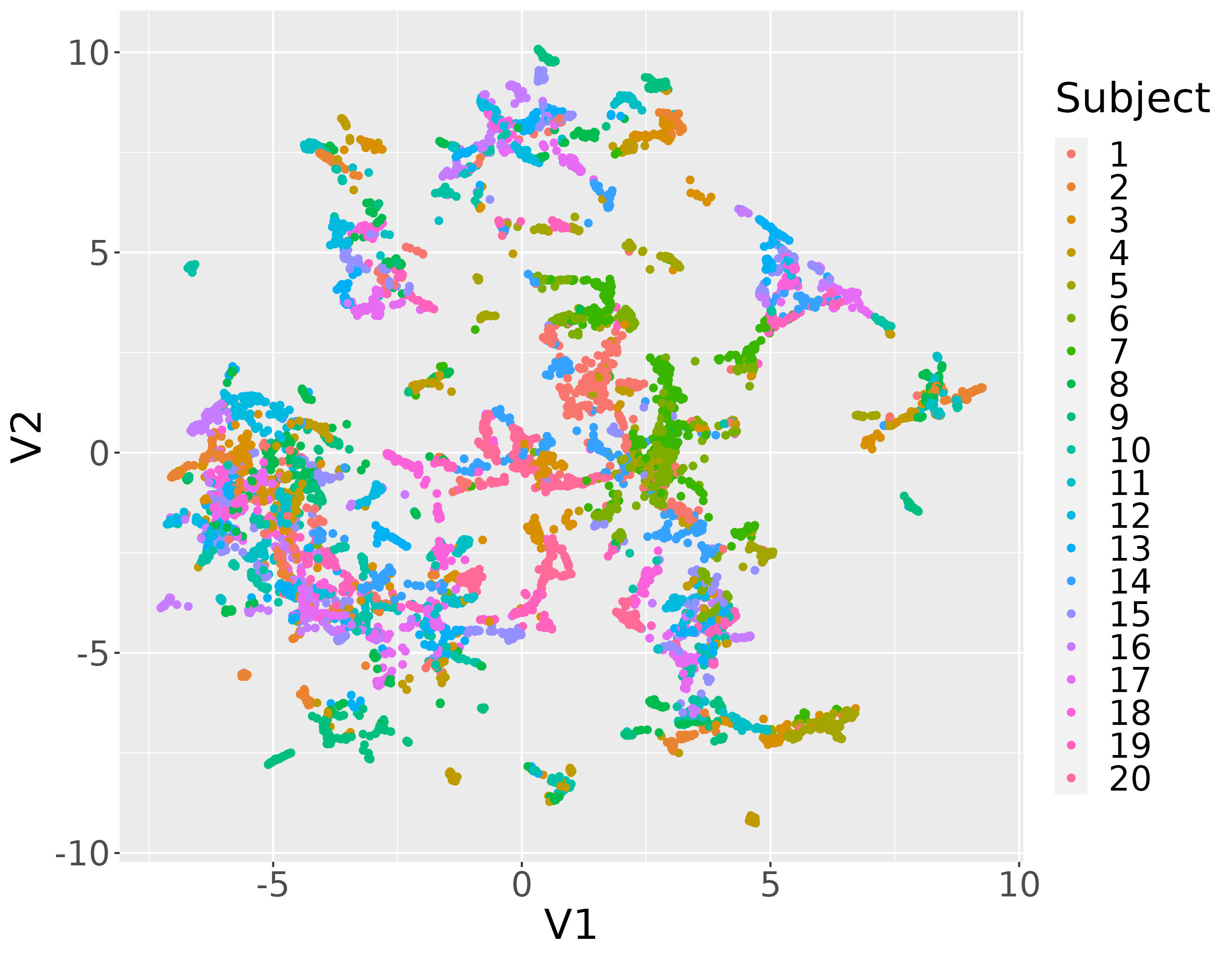}}
  \\
  \subfloat[Surface Vowel\label{fig:umap_sub_vowel}]{\includegraphics[width=0.3\linewidth]{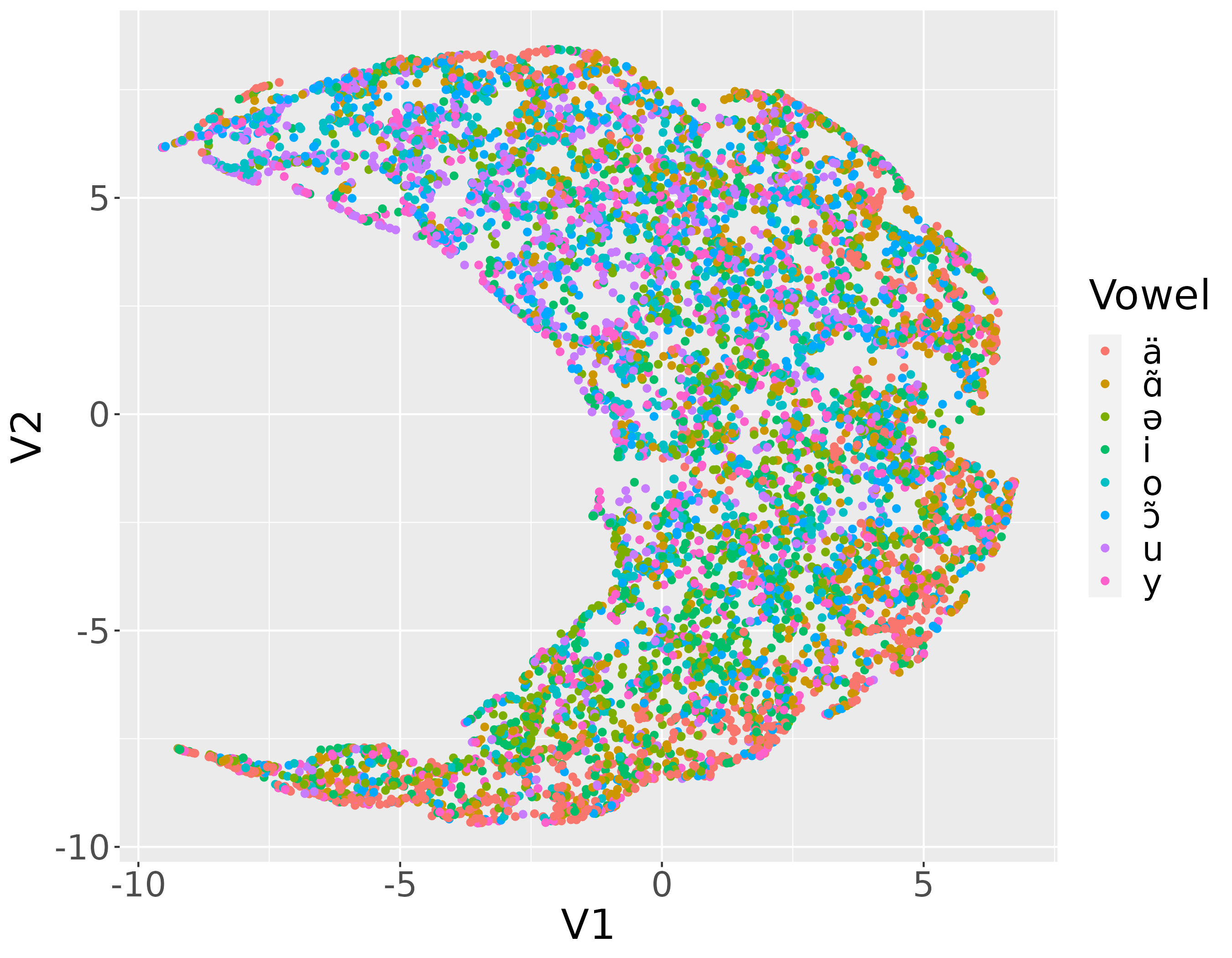}}
  \hfill
  \subfloat[Surface Sex\label{fig:umap_sub_sex}]{\includegraphics[width=0.3\linewidth]{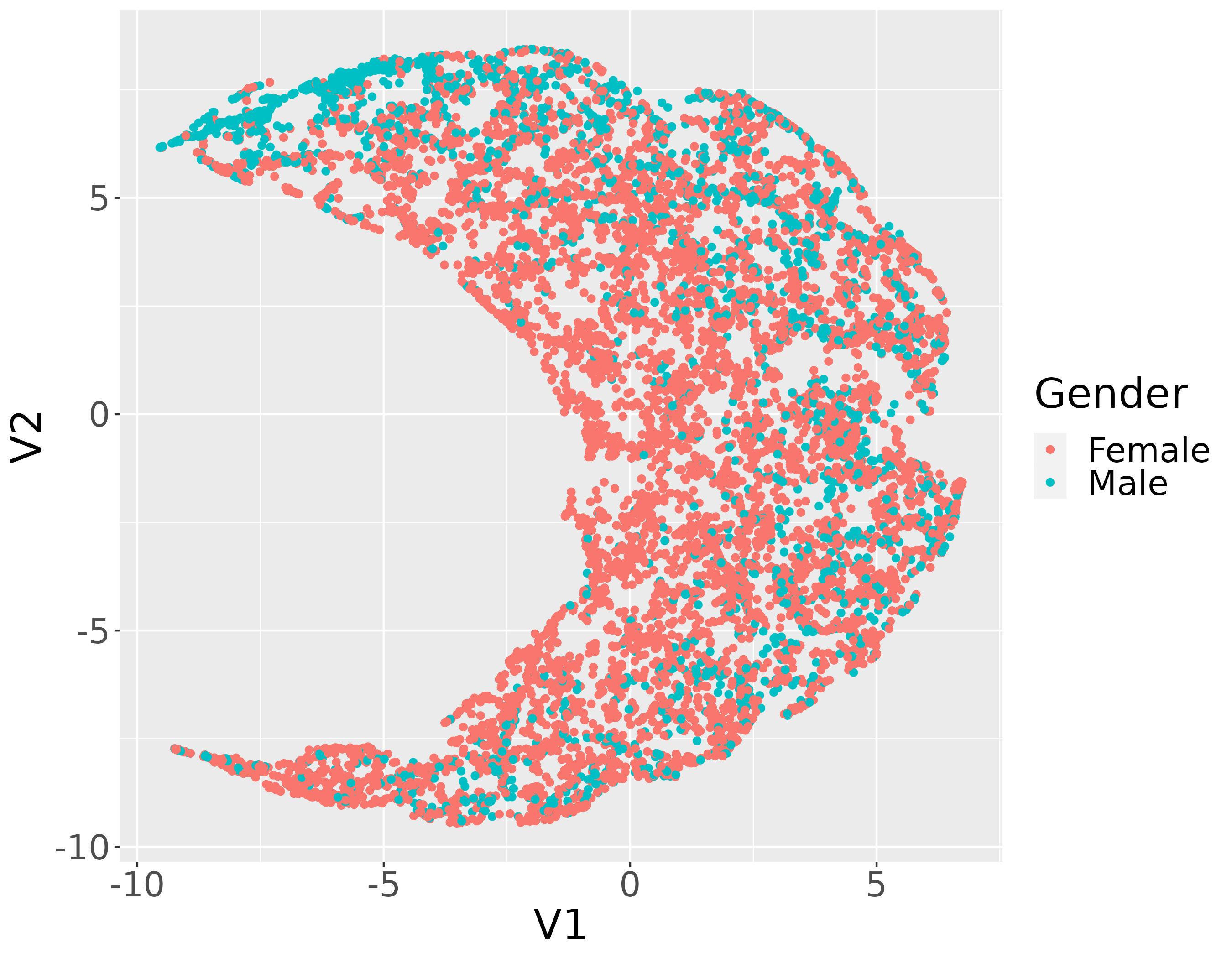}}
  \hfill
  \subfloat[Surface Subject\label{fig:umap_sub_subject}]{\includegraphics[width=0.3\linewidth]{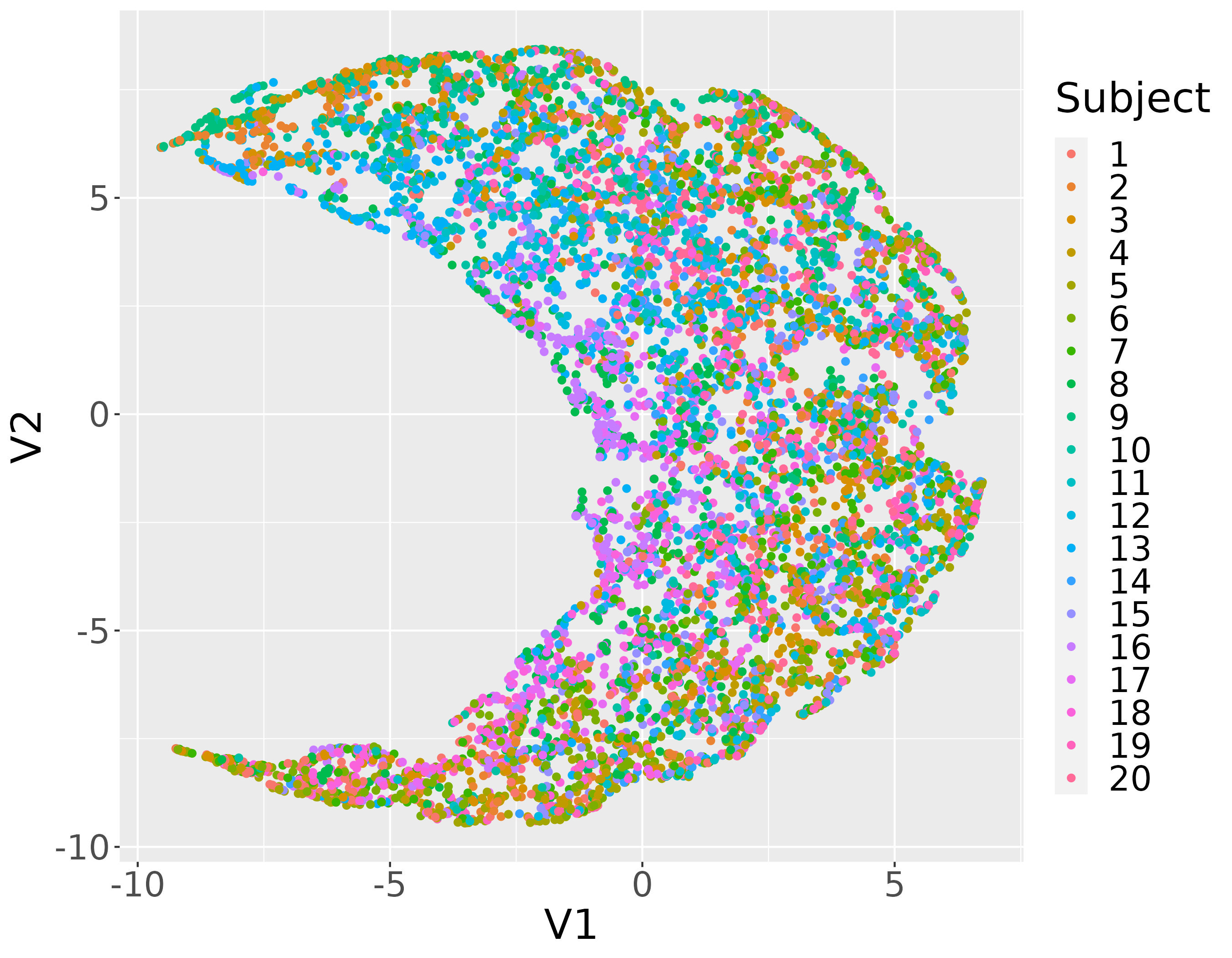}}
  \\
  \subfloat[Zeros Vowel\label{fig:umap_zeros_vowel}]{\includegraphics[width=0.3\linewidth]{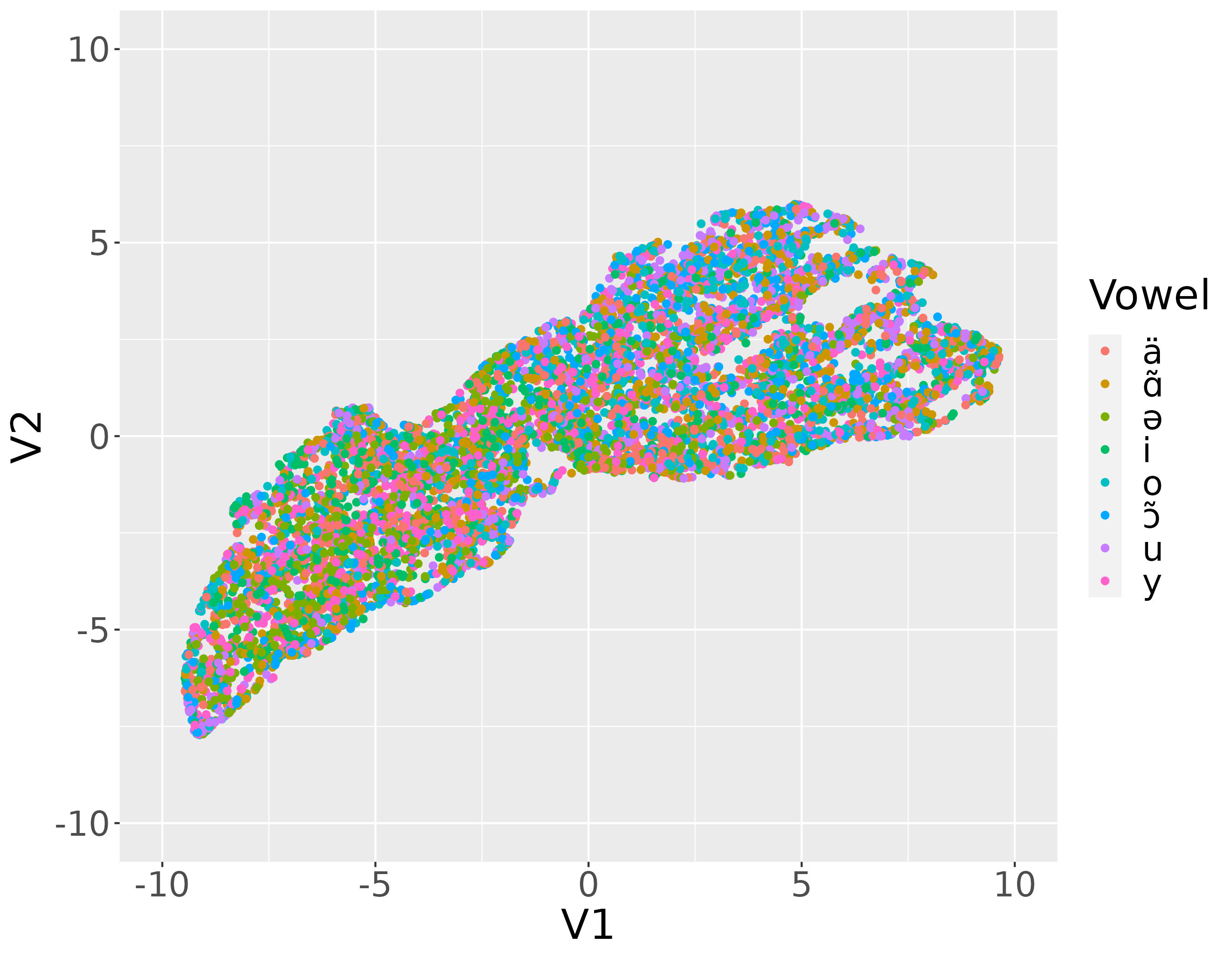}}
  \hfill
  \subfloat[Zeros Sex\label{fig:umap_zeros_sex}]{\includegraphics[width=0.3\linewidth]{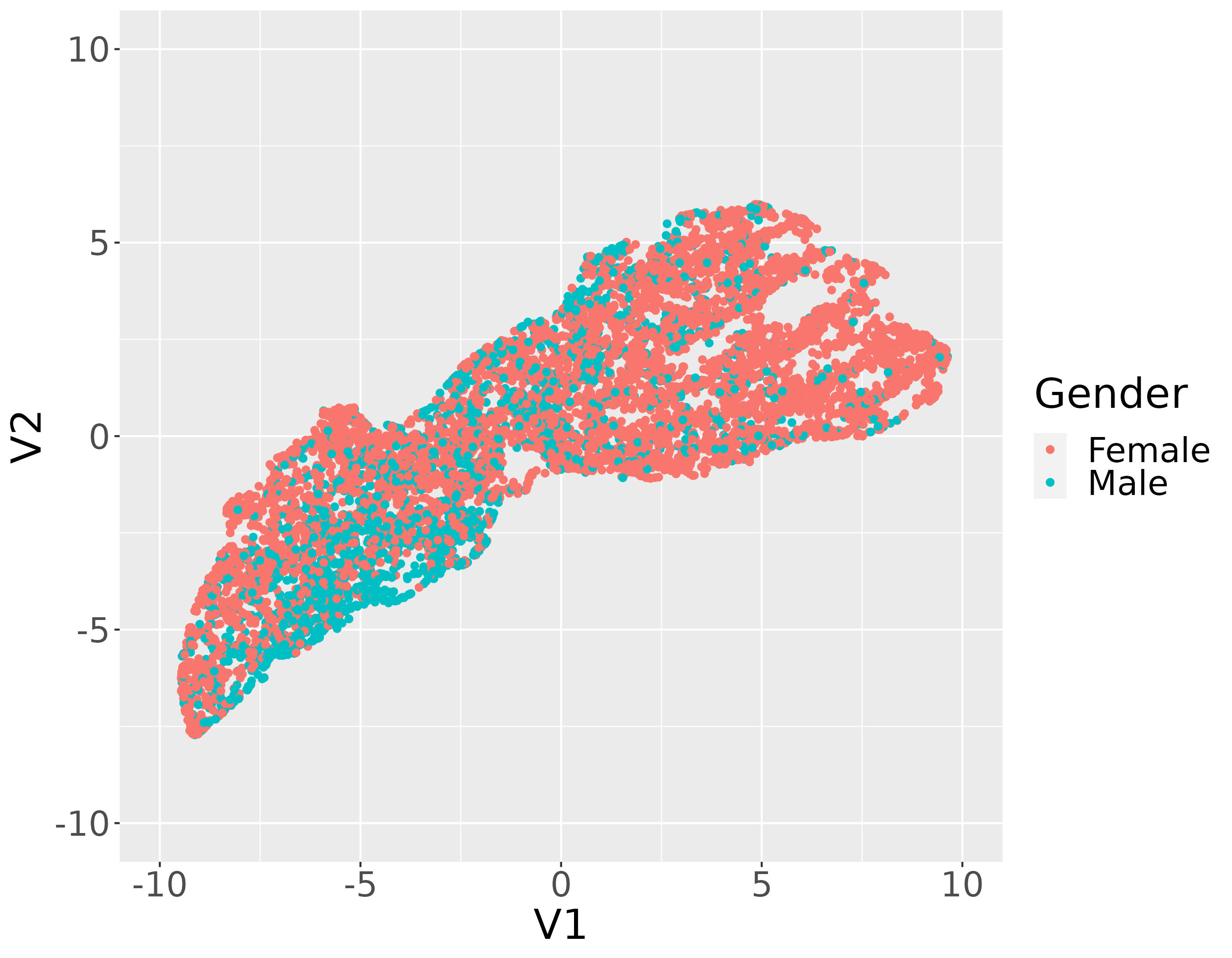}}
  \hfill
  \subfloat[Zeros Subject\label{fig:umap_zeros_subject}]{\includegraphics[width=0.3\linewidth]{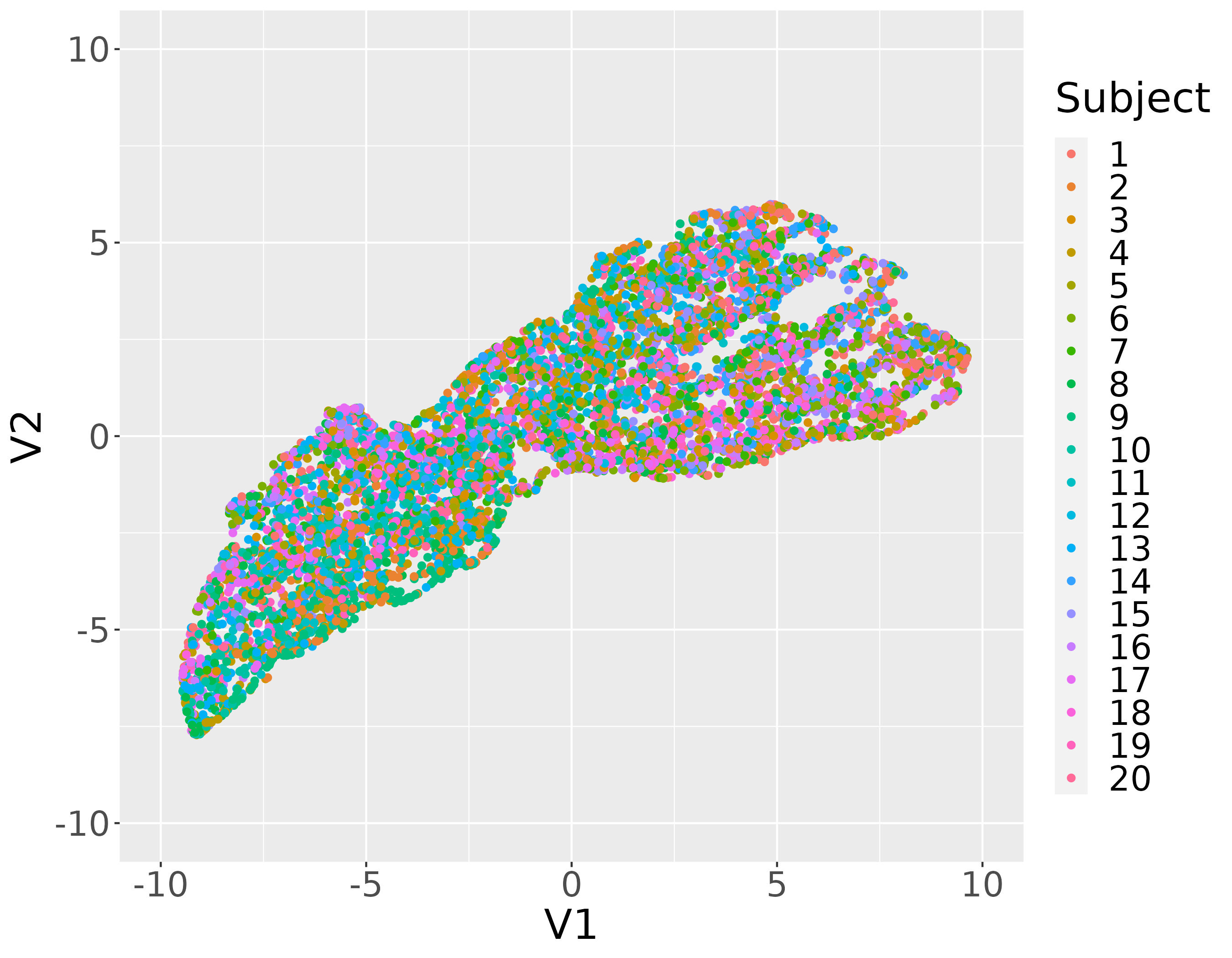}}
  \\
  \subfloat[Takens Vowel\label{fig:umap_taken_vowel}]{\includegraphics[width=0.3\linewidth]{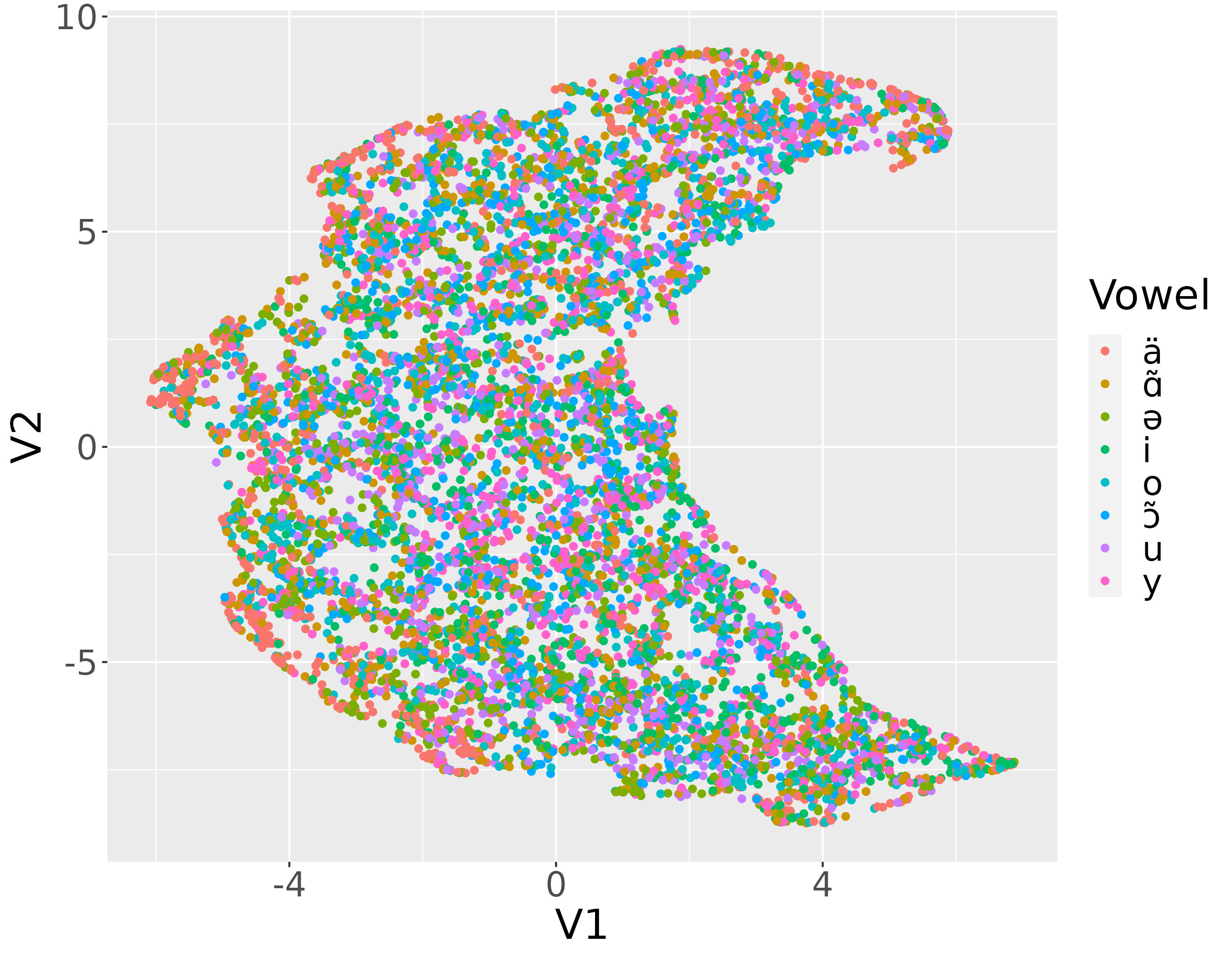}}
  \hfill
  \subfloat[Takens Sex\label{fig:umap_taken_sex}]{\includegraphics[width=0.3\linewidth]{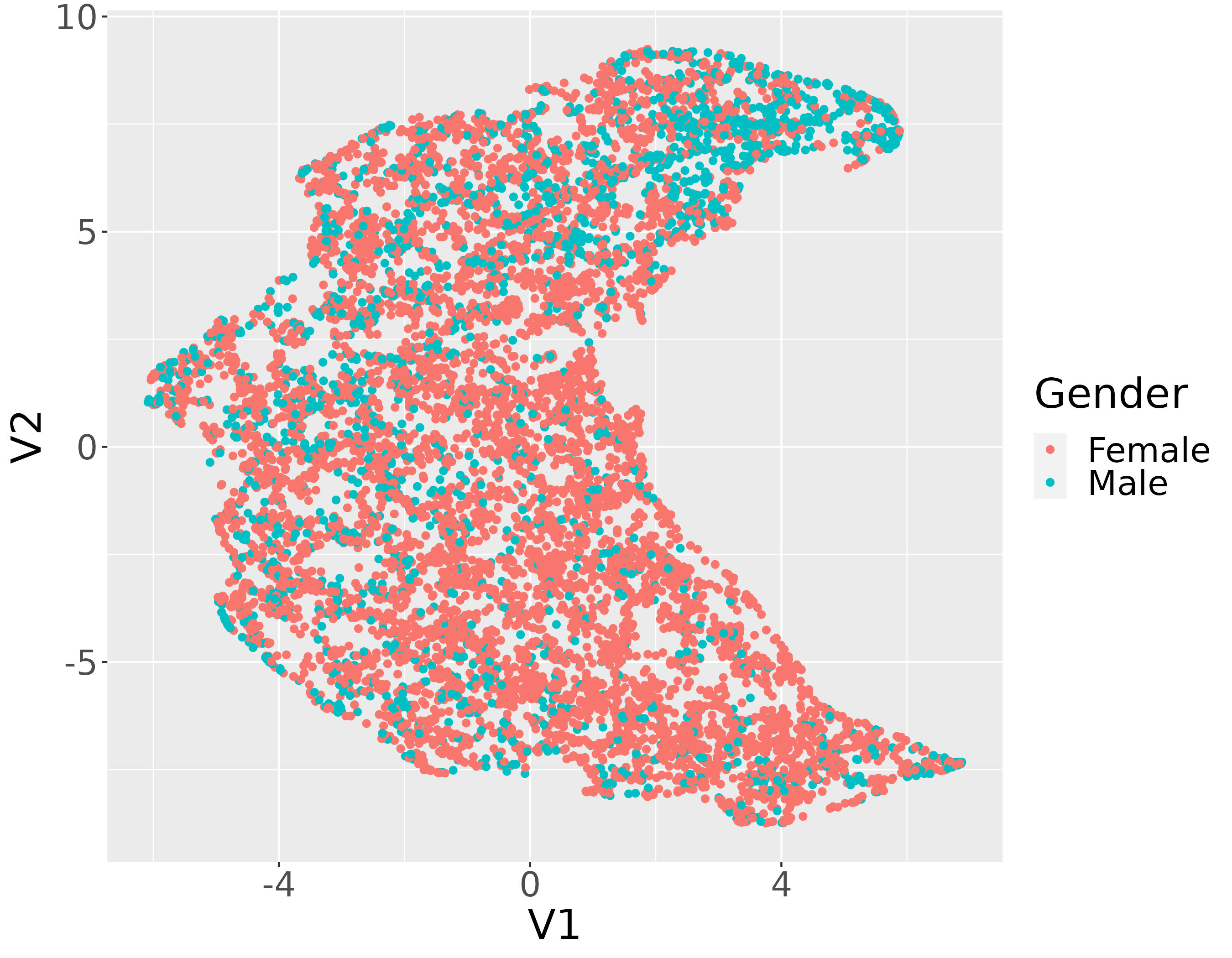}}
  \hfill
  \subfloat[Takens Subject\label{fig:umap_taken_subject}]{\includegraphics[width=0.3\linewidth]{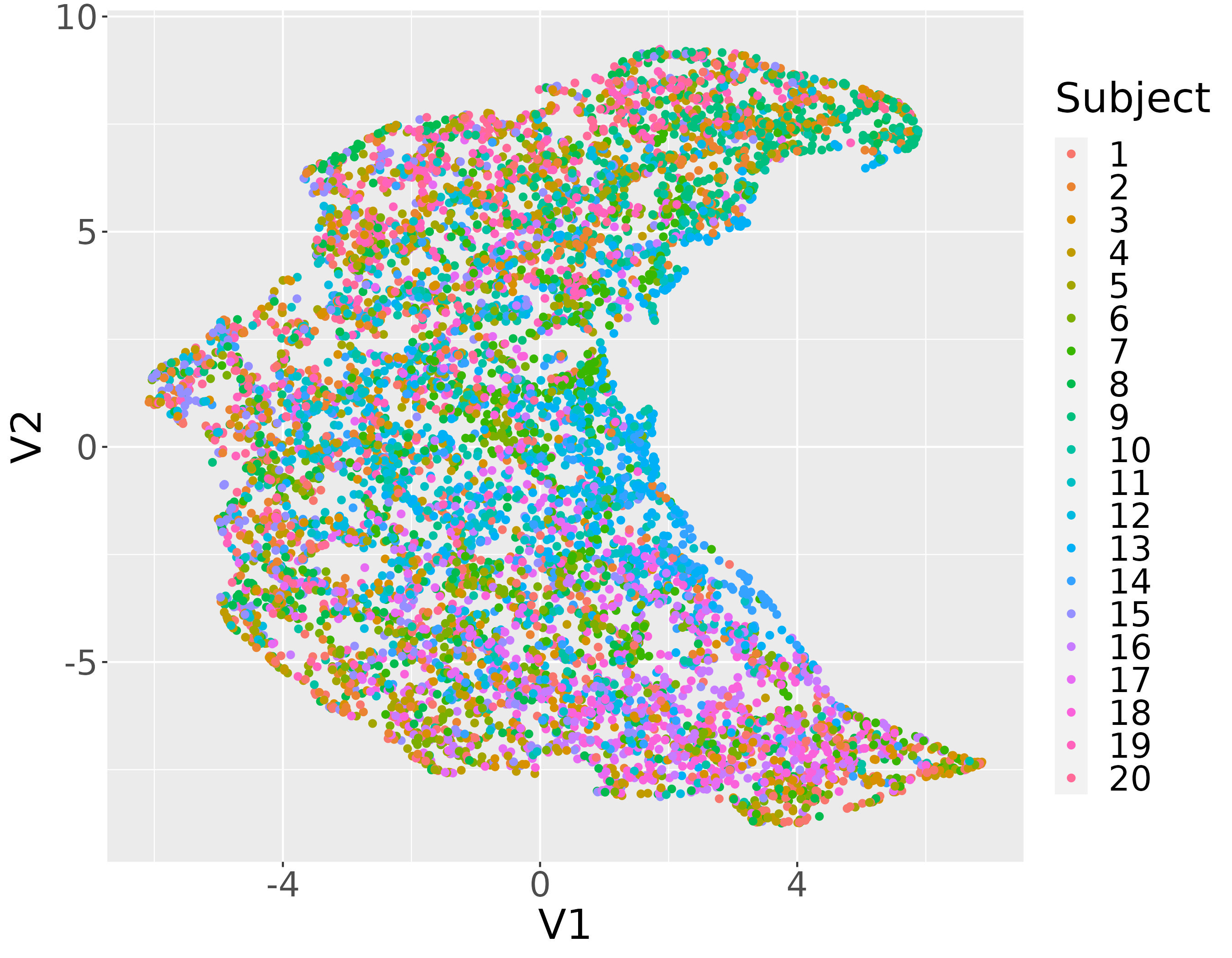}}
  \caption{Projection of the records on the 2d space learned with UMAP, depending on the input of the algorithm: MFCC, topological variables from the spectrogram's surface, the spectrogram's zeros or the Taken's embedding.}
  \label{fig:umap} 
\end{figure*}

In order to describe better the behavior of the homological features associated to the different representation spaces, we provide a visual and qualitative help using the Uniform Manifold Approximation and Projection for Dimension Reduction (UMAP) algorithm. Figure~\ref{fig:umap} represents the projection of the recordings onto the 2d space learned with the UMAP. It distinguishes MFCC, persistent variables accordingly to the representation space, and the three problems considered therein.


Figure~\ref{fig:umap} reveals that the manifold learned on the MFCCs creates clear clusters for vowels. This is less the case for gender, where observations from both classes are quite close in this space. Same comment applies for individuals. Although some observations are centered, the distribution of individuals in this space can be quite spread out. 
It is even harder to see a pattern for the manifold learned on persistent variables, whatever the representation space. In fact, the manifolds learned from the persistent variables are much more connected, and we do not distinguish clear clusters in these spaces. 



Interestingly, the overall shapes of manifolds learned on Taken's embeddings and on spectrogram's surface look quite similar, and rotated by 180°. While connectivity is still present, manifolds learned on topological variables extracted from zero's spectrogram looks very different. This might be explained by the number of coordinates in the representation space within the persistent homology. Taken's embedding and the spectrogram's surface lie in a 3-dimensional space, whereas the spectrogram's zeros are in the plane.  

\section{Discussion}

The main results that emerge from the present study are fourfold. Firstly, topological information improved some classification results. Secondly, the choice of the signal representation space has an impact on the topological information that is extracted. Thirdly, the best way to use a persistence diagram for a classification task depends on the problem. Finally, topological information extracted from the spectrogram's zeroes appeared to be particularly complementary to the MFCC and provided the best results out of all tested models for individual prediction.

\subsection{On the improvements on the prediction of labels}

On the one hand, no matter how the information is extracted from the persistence diagrams, topological information never outperformed the MFCCs. On the other hand, adding topological variables to frequency variables improves the results in two of the three considered classification problems. We thus confirmed results already described in the literature  \cite{carriereStableTopologicalSignatures2015, xuTopologicalDataAnalysis2021,chazalIntroductionTopologicalData2021}. Topological descriptors of signals bring complementary information to MFCCs, which are specifically designed for human speech analysis. 

These results contribute to fill an existing gap in the analysis of time-varying data, as noted in some recent reviews \cite{henselSurveyTopologicalMachine2021}. They are encouraging enough to continue investigating the performance of topologically-augmented machine learning approaches even for natural signal suffering from much lower signal-to-noise ratio as one might consider robust filtration alike Distance-To-Measure introduced in \cite{ChazalDTM2014}.


Our results show no improvement for the gender classification problem. Remark that it is the simplest problem as it is binary classification, while the other have 8 and 20 classes, for vowels and individuals prediction, respectively. Another possible reason is the pitch difference between  men and women. Topological characteristics are certainly invariant to pitch modulations, and MFCC's particularly well suited.

When the problem becomes more complicated, topological information proved to be useful. The most notable improvement is for the prediction of the speaker. Persistent homologies seem to carry this additional information needed to improve the classification performance.

\subsection{Different objects, different topologies}

\subsubsection{The difference between representation spaces}

As already noticed, it is difficult to identify one representation of the signal as being better than another. Nevertheless, there are two points to bear in mind from the classification results according to the initial representation space:

\begin{itemize}
    \item For the two problems where topological information improves the results, i.e., individual and vowel classification, the persistent homologies extracted from the time-frequency plane of the spectrogram zeros improve the OOB error the most. Taken's embeddings are the representation with the least improvement for both problems. Thus, access to higher dimensional homologies does not seem to provide discriminative information about the signal.
    \item The aggregation of topological information is also advantageous. It allows the best improvement in the prediction of vowels and, when it does not allow the best improvement in the prediction of individuals, the improvement is better than for the other two representations. Moreover, for the models trained only on the persistent variables, without the MFCCs, the best results are always obtained when all the representations are combined. The homological information of the representation spaces therefore seems complementary.
\end{itemize}

This shows that a particular representation space, reveal different salient topological features. Very often, Taken's embeddings are considered in topological signal analysis. In this paper, we introduced two ways to derive topological information from spectrogram that have the advantage of being more interpretable. 

\subsubsection{A different perspective}


The topological signatures we computed on the different signal representation spaces are complementary to more classical descriptors such as MFCCs. This has been highlighted in supervised classification problems \cite{carriereStableTopologicalSignatures2015, xuTopologicalDataAnalysis2021, chazalIntroductionTopologicalData2021}.

The manifolds we learn from the persistent variables, presented in Figure~\ref{fig:umap}, are strongly connected. The topological approach is metric-free and based on the connectivity \cite{carlssonTopologyData2009}. A topological latent space does not clearly cluster the data. 

We find a similar observation in the topological autoencoder \cite{moorTopologicalAutoencoders2020}. This property makes entangled structures appear that are impossible in a clear spatial separation. This would be useful to illustrate a hierarchy in the data, with a parent-child structure. 

The value of the topological approach depends on the problem \cite{wassermanTopologicalDataAnalysis2018} and is particularly suited to problems requiring analysis of nested categories. If one expects such a structure in the data, and we wish to highlight it, it is interesting to examine its topology, in order to look at the problem from a blind spot to more classical analysis tools.

\subsection{On the more present persistent variables}

Finally, regarding the vectorization of persistence diagrams, it seems difficult to characterize topological signatures, at least among those tested. We confirm a result already present in other reviews studying this question \cite{barnesComparativeStudyMachine2021}, which find that the most interesting topological signatures seem to depend on the studied data set. Here we found that mapping the persistence diagram onto a functional space, such as \cite{chazalStochasticConvergencePersistence2014,adamsPersistenceImagesStable2017}, was less efficient than taking a set of scalars that summarize the information contained in the diagram. Nevertheless, these approaches have the advantage of having well established stability properties and have been used elsewhere \cite{bubenikPersistenceLandscapeIts2020a, liuApplyingTopologicalPersistence2016, kimPLLayEfficientTopological2020}. This does not disqualify them for other problems, and they might even be effective on this problem, if handled more carefully (for example, by tuning finely the different parameters they depend on, or by computing well-chosen summary data).

This question also raises the issue of what is considered to be topological noise. Topological noise generally designates homological features with very short lifespans. It can be seen on persistent diagrams as the set of points along the main diagonal. We might ask ourselves, is topological noise really noise? Indeed, it has already been noticed that topological noise can be useful to characterize a signal \cite{patrangenaruChallengesTopologicalObject2019}. The persistence diagrams obtained after alpha-complex filtration on the zeros of the spectrograms are visually the diagrams with the most persistent homologies, many of them close to the diagonal, as can be seen in Figure~\ref{fig:diagrams}. Yet, it is this representation that gives the best improvements. Perhaps these homologies are not just noise and contain information. It is important to bear this in mind when choosing a way to use the information contained in persistence diagrams, as not all methods treat elements close to the diagonal in the same way.

Finally, we did not test all the existing approaches to this question. In fact, we adopted a strategy consisting of extracting information from the persistence diagrams, either by mapping persistence diagrams onto function space (persistent landscape lives in Banach space) or by calculating vectors to describe them. 

\section{Conclusion}

We discussed the potential added value of the topological approach to sound signal processing by studying the differences according to the representation space of the signal. We tested it on three classification problems, predicting the gender of the speaker, the pronounced vowel and the identity of the speaker. For two of these problems, vowel and identity prediction, the topological features improve the results compared to the baseline. Although it is difficult to distinguish one representation space as being more informative than another, it seems that the topological descriptors computed on each of them are complementary. 
Our results suggest the use of less common representations than Taken's embeddings, such as spectrogram's surfaces or spectrogram's zeros. For individual classification, the zeros give the best results with the topologically-augmented approach. Moreover, parameter selection and interpretation of a spectrogram is simpler than Taken's embeddings.
We analyzed different ways of vectorizing information from persistence diagrams. The best results were obtained using  a set of persistent variables regardless of the representation space. This shows that the topological approach offers a complementary and interesting angle of analysis. 
It is not sufficient on its own to discriminate the signal, but it does provide additional information about its hierarchical structure. It would be interesting to analyze more theoretically the complementarity of the representation spaces of the signal, and to see the possibility of working directly in the space of persistence diagrams.

In the end, we believe that TDA has a lot to offer to sound signal processing, in particular for individual recognition problems. 




\vspace{11pt}


\vfill

\bibliographystyle{ieeetr}
\bibliography{topological_data_analysis}

\end{document}